\documentclass[floatfix,amssymb,prl,twocolumn,superscriptaddress,preprintnumbers,nofootinbib, aps]{revtex4-1}
		
\usepackage{amssymb,amsmath,verbatim,mathtools,needspace,enumitem,etoolbox,graphicx,physics,microtype,afterpage,bm,soul}
\usepackage[dvipsnames]{xcolor}
\definecolor{linkcolor}{rgb}{0.0,0.3,0.5}
\usepackage[unicode, colorlinks=true, linkcolor=linkcolor, citecolor=linkcolor, filecolor=linkcolor,urlcolor=linkcolor, pdfusetitle]{hyperref}
\usepackage{orcidlink}
\usepackage[all]{hypcap}
\usepackage[T1]{fontenc}
\usepackage[utf8]{inputenc}
\usepackage{tabularx}
\usepackage{float}
\usepackage{placeins}
\usepackage{lmodern}
\usepackage{ragged2e}
\allowdisplaybreaks
\usepackage{tikz}
\usepackage{color}
\usepackage{framed}
\usepackage{hyperref}
\hypersetup{colorlinks, citecolor=darkblue, linkcolor=black, urlcolor=darkblue}
\definecolor{rossos}{cmyk}{0,1,1,0.55}
\definecolor{bluscuro}{rgb}{0.15, 0.2, .85}
\definecolor{bluchiaro}{cmyk}{1,.3,0.,0.1}
\definecolor{ForestGreen}{rgb}{0.13, 0.55, 0.13}
\definecolor{darkblue}{rgb}{0,0,0.4}

\newcommand{\sig}{{\texttt{NSBHc}}}   
\newcommand{\pp}{{0}}       

\newcommand{\msun}{M_{\odot}}
\newcommand{\be}{\begin{equation}}
\newcommand{\ee}{\end{equation}}

\def\BH{\text{\tiny BH}}

\newcommand{\NS}{\text{\tiny NS}}

\newcommand{\GW}{\text{\tiny GW}}

\def\lsim{\mathrel{\rlap{\lower4pt\hbox{\hskip0.5pt$\sim$}}
    \raise1pt\hbox{$<$}}}         
\def\gsim{\mathrel{\rlap{\lower4pt\hbox{\hskip0.5pt$\sim$}}
    \raise1pt\hbox{$>$}}}         

\newcommand{\jhu}{William H.\ Miller III Department of Physics and Astronomy, Johns Hopkins University, \\ 3400 North Charles Street, Baltimore, Maryland, 21218, USA}
\newcommand{\columbia}{Department of Physics and Columbia Astrophysics Laboratory, Columbia University, New York, NY 10027, USA}

\begin{document}

\title{
Subsolar neutron star-black hole binaries from collapsars:\\Gravitational waveform models and detectability in current searches
}

\author{Vishal Baibhav\orcidlink{0000-0002-2536-7752}}
\email{vb2630@columbia.edu}\thanks{NASA Einstein Fellow}
\affiliation{\columbia}

\author{Valerio De Luca\orcidlink{0000-0002-1444-5372}}
\email{vdeluca2@jh.edu}
\affiliation{\jhu}

\author{Loris Del Grosso\orcidlink{0000-0002-6722-4629}}
\email{ldelgro1@jh.edu}
\affiliation{\jhu}

\author{Emanuele Berti\orcidlink{0000-0003-0751-5130}}
\email{berti@jhu.edu}
\affiliation{\jhu}

\author{Brian D.~Metzger\orcidlink{0000-0002-4670-7509}}
\email{bdm2129@columbia.edu}
\affiliation{\columbia}
\affiliation{Center for Computational Astrophysics, Flatiron Institute, 162 5th Avenue, New York, NY 10010, USA}

\author{Lam Hui\orcidlink{0000-0001-7003-4132}}
\email{lh399@columbia.edu}
\affiliation{\columbia}

\begin{abstract}
\medskip
\noindent
Detecting subsolar objects in binary mergers would be an exceptional discovery in gravitational wave astronomy.  So far, dedicated searches for subsolar coalescences reported no confident detections, setting upper bounds on their merger rate. 
Given that standard stellar evolution is not able to produce subsolar candidates, a confident detection would point either to new physics or to unconventional astrophysical pathways. These include collapsars, where the outer parts of neutrino-cooled accretion disks around newborn stellar-mass black holes are subject to fragmentation and give rise to subsolar neutron stars, which can later merge with the central black hole. We show that current subsolar searches may miss a significant fraction of these neutron star–black hole mergers, because their templates do not account for the small mass ratios, the tidal disruption of the neutron star, and the effects of the surrounding gas. This result motivates improving the waveform modeling used in current subsolar searches and building adequate template banks tailored to these systems.
\end{abstract}


\maketitle


\vspace{0.1cm}
\noindent{{\bf{\em Introduction.}}} 
The detection of a compact object with mass below 
$\sim1\,\msun$ 
would be one of the most consequential outcomes of the current gravitational-wave (GW) observational program~\cite{LIGOScientific:2016vlm,LIGOScientific:2018mvr,LIGOScientific:2020ibl,LIGOScientific:2021usb,KAGRA:2021vkt,LIGOScientific:2025hdt,LIGOScientific:2025slb,LIGOScientific:2026wfs}. Standard stellar evolution produces neither remnant black holes (BHs) nor ordinary core-collapse neutron stars (NSs) in this range~\cite{Chandrasekhar:1931ih,Shapiro:1983du,Burrows:2020qrp,Lattimer:2021emm,Janka:2025tvf,Karim:2026bva}, so a confident subsolar detection would point either to new physics or to an unconventional formation pathway~\cite{Shandera:2018xkn,Singh:2020wiq,Essick:2024olf,Krnjaic:2026sgl,Gao:2026zwc}. Dedicated searches for subsolar coalescences have been carried out by the LIGO-Virgo-KAGRA (LVK) Collaboration and independent groups across all observing runs~\cite{Abbott:2018oah,Authors:2019qbw,Nitz:2020bdb,Nitz:2021mzz,Nitz:2022ltl,LIGOScientific:2021job,Phukon:2021cus,LVK:2022ydq,Kacanja:2024hme,Soni:2024cdb,Kacanja:2026byy,LIGOScientific:2026wxz,LIGOScientific:2026kef,Kacanja:2026nwi}. The most recent ones, including the LVK search on data from the second part of the fourth observing run (O4b), report no confident detection and bound the merger rate to $\sim(50-5000)\,{\rm Gpc^{-3}\,yr^{-1}}$ (90\% confidence), depending on the component masses~\cite{LIGOScientific:2026kef,Kacanja:2026nwi}. However, two low-significance candidates, S250818k and S251112cm~\cite{2025GCN.41437....1L,ligo_scientific_collaboration_ligovirgokagra_2025}, have triggered extensive electromagnetic follow-ups~\cite{Kasliwal:2025keb,Franz:2025uan,Gillanders:2025fwf,Yang:2025dtj,Ackley:2026zlf,Vieira:2026eof,ODwyer:2026caq,Hall:2026gov,Tejera:2026dpr,Paek:2026kat,Liu:2026hmk}.

These searches have been primarily motivated by primordial BHs (PBHs)~\cite{Zeldovich:1967lct,Hawking:1971ei,Carr:1974nx,Carr:1975qj} (see~\cite{Sasaki:2018dmp,Carr:2020gox,Green:2020jor,LISACosmologyWorkingGroup:2023njw,Byrnes:2025tji} for reviews), which can form subsolar binaries in the early Universe~\cite{Cholis:2016kqi,Bird:2016dcv,Clesse:2016vqa,Sasaki:2016jop,Raidal:2018bbj,Raidal:2024bmm} and are now limited below $1\%$ of the dark matter in the $(0.1-1)\,\msun$ window~\cite{Carr:2020gox,Green:2020jor,Andres-Carcasona:2024wqk,Carr:2026hot,LIGOScientific:2026kef,Kacanja:2026nwi}. Since PBH binaries are well modeled as point masses in vacuum, current subsolar searches use quasicircular, aligned-spin template waveforms with mass ratios larger than $0.1$~\cite{LIGOScientific:2021job,LVK:2022ydq,LIGOScientific:2026wxz,LIGOScientific:2026kef}, without environmental effects. Astrophysical subsolar sources, whose waveforms carry imprints of finite size and of the environment that also set them apart from PBHs~\cite{Cardoso:2017cfl,Cardoso:2019rvt,Cardoso:2019upw,Barsanti:2021ydd,Franciolini:2021tla,Franciolini:2021xbq,Cole:2022fir,Coogan:2021uqv,Crescimbeni:2024cwh,Golomb:2024mmt,Crescimbeni:2024qrq,DeLuca:2024uju,Russo:2025azm,DeLuca:2025bph,Begnoni:2025aqc,Corman:2026lbt}, may therefore be missed. For subsolar NS binaries, tidal dephasing and termination at Roche-lobe overflow already reduce the sensitive volume of such template banks substantially~\cite{Bandopadhyay:2022tbi}.

The best-motivated such sources are NSs born in the outer regions of collapsar disks~\cite{Piro:2006ja,Wu:2023qeh,Metzger:2024ujc,Lerner:2025dkd,Chen:2025uwd}. There, the neutrino-cooled accretion disk around the BH formed in the collapse of a rapidly rotating massive star, believed to be the engine of long gamma-ray bursts (LGRBs)~\cite{MacFadyen:1998vz,Woosley:2006fn} and broad-lined Type Ic supernovae~\cite{Metzger:2024ujc}, fragments into neutron-rich clumps that form NSs of $\sim(0.01-1)\,\msun$~\cite{Lattimer:2004pg,Metzger:2024ujc,Chen:2025uwd}. These produce two classes of GW sources~\cite{Metzger:2024ujc,Baibhav:2026iny}: binary NS (BNS) mergers orbiting the central BH as hierarchical triples, and the inspiral of a surviving NS into the central BH (a NS-BH system). In two companion papers, we characterized the waveform signatures of both channels~\cite{Baibhav:2026iny} and their stochastic background~\cite{DeLuca:2026eis}, whose non-detection in O4 already bounds the fraction of fragmenting collapsars~\cite{DeLuca:2026eis,Virgo:2025aai}. A dedicated search for resolvable events could tighten this bound~\cite{DeLuca:2026eis}, but only if gravitational waveform templates include these sources. 

In this Letter, we focus on NS-BH mergers (the loudest merger channel in collapsars~\cite{Baibhav:2026iny,DeLuca:2026eis}) and ask whether current searches would detect or miss them. The signal lies largely outside the parameter space covered by existing template banks: its intermediate mass ratios $\sim10^{-3}-10^{-1}$ extend below the lower limit $0.1$ of the LVK bank~\cite{DeLuca:2026eis,Cheung:2025grp} (see~\cite{RamisVidal:2026ycb,Markin:2026eyc} for subsolar NS-BH mergers with comparable masses); the inspiral may end in tidal disruption of the NS well below the innermost stable circular orbit (ISCO) frequency, up to which templates extend; and the phase carries a large tidal deformability, as well as a spin-induced quadrupole moment hundreds of times the Kerr value, residual eccentricity~\cite{Lenon:2020oza,Dhurkunde:2023qoe,Morras:2025xfu,Planas:2025plq,Kacanja:2025kpr}, and negative post-Newtonian (PN) dephasing from gas-driven migration, none of which are included in the template waveforms of current searches. We model the signal with a frequency-domain waveform that includes all these effects and compute its fitting factor against the template family of the LVK O4b search. 
This quantifies the fraction of collapsar events that current searches would miss,  gives approximate bounds on the collapsars rate, and provides a waveform model for dedicated searches, whose triggers could be cross-matched with the LGRB and supernova accompanying each merger~\cite{Baibhav:2026zuy}. Henceforth, we adopt geometrical units ($G = c = 1$).

\vspace{0.1cm}
\noindent{{\bf{\em A primer on the collapsar model.}}}
Core-collapse NSs inherit the Chandrasekhar mass of the iron core, given by $M_\text{\tiny Ch}\simeq1.45\,\msun\,(Y_e/0.5)^{2}$ in terms of the electron fraction $Y_e$, with resulting masses larger than $1.2\,\msun$~\cite{Chandrasekhar:1931ih,Shapiro:1983du,Sukhbold:2015wba,Burrows:2019rtd,Ertl:2019zks,Woosley:2020mze,Burrows:2020qrp,Lattimer:2021emm,Janka:2025tvf,Karim:2026bva}. 
Lighter NSs require neutron-rich matter characterized by $Y_e < 0.5$, a condition realized in collapsars~\cite{Lattimer:2004pg,Piro:2006ja,Wu:2023qeh,Metzger:2024ujc,Lerner:2025dkd,Chen:2025uwd}, where a BH of mass $M_\BH\simeq(3 - 30)\,\msun$~\cite{Schneider:2020vvh} is surrounded by a hot, dense disk accreting at $\dot M_\BH\sim(0.1 - 1)\,\msun\,{\rm s}^{-1}$~\cite{MacFadyen:1998vz,Woosley:2006fn,Gottlieb:2023cgm}.
For a disk aspect ratio $\mathbb{H}=h/r$, the gravitationally unstable outer regions can fragment if cooling is sufficiently rapid, producing clumps of mass~\cite{Chen:2025uwd}
\begin{equation}
m_\NS\simeq\mathbb{H}^3M_\BH\simeq0.14\,\msun
    \left(\frac{\mathbb{H}}{0.3}\right)^3
    \left(\frac{M_\BH}{5\,\msun}\right)\,.
    \label{eq:clumpmass}
\end{equation}
The resulting NS-BH mass ratio, $q\equiv m_\NS/M_\BH\simeq\mathbb{H}^3$, is independent of $M_\BH$ at fixed $\mathbb{H}$. Although multiple fragments may form, interact, scatter, and merge~\cite{Baibhav:2026zuy}, we assume for simplicity that a single NS-BH binary forms.
In the absence of migration due to gas, the time to coalescence is~\cite{Peters:1963ux,Peters:1964zz}
\begin{align}
    \tau_\GW\simeq{}&12\,{\rm min}
    \left(\frac{m_\NS}{0.1\,\msun}\right)^{-1} \left(\frac{M_\BH}{5\,\msun}\right)^{-2/3}
    \left(\frac{f_{\text{\tiny NS-BH},i}}{20\,{\rm Hz}}\right)^{-8/3}\,,
    \label{eq:tauGW}
\end{align}
for which the in-band inspiral accumulates $\mathcal{O}(10^4-10^5)$ cycles in ground-based detectors~\cite{Baibhav:2026iny}.

For the internal composition of the NSs, we adopt for simplicity an ideal gas of non-relativistic degenerate neutrons, i.e., an $n=3/2$ polytropic equation of state (EoS)~\cite{Shapiro:1983du,Metzger:2024ujc}, for which subsolar NSs are large, $R_\NS\simeq34\,{\rm km}\,(m_\NS/0.1\,\msun)^{-1/3}$~\cite{Baibhav:2026iny}. More realistic nuclear-matter models admit stable NSs only above $\simeq0.1\,M_\odot$~\cite{1991ApJ...369..422C}, while self-bound strange-quark-matter stars can extend to arbitrarily small masses~\cite{Gondek:1997fd,Haensel:1986qb,Prakash:1995uw,Prakash:1996xs}; the former are comparably large near this floor, whereas the latter are more compact~\cite{Baibhav:2026iny}.

A binary born at $a_{\text{\tiny NS-BH},i}$ initially has GW frequency
\begin{equation}
    \label{eq:fNSBHi}
    f_{\text{\tiny NS-BH},i} 
    \simeq 13\,{\rm Hz}\,\left(\frac{5\,\msun}{M_\BH}\right)\left(\frac{a_{\text{\tiny NS-BH},i}}{100\,R_g}\right)^{-3/2}\,,
\end{equation}
with $R_g=M_\BH$ the gravitational radius of the BH. The NS then orbits until it is either disrupted or reaches the ISCO of the BH, whichever comes first. The BH shreds the NS at the Roche separation $2R_\NS(M_\BH/m_\NS)^{1/3}$~\cite{Shapiro:1983du,Vallisneri:1999nq,GomezLopez:2025aiy}, i.e., at the GW frequency $f_\text{\tiny Roche}\propto m_\NS^{1/2}R_\NS^{-3/2}$, which for the polytropic EoS is given below, together with the ISCO frequency of a non-spinning BH~\cite{Bardeen:1972fi}:
\begin{equation}
    f_\text{\tiny Roche} \simeq 65\,{\rm Hz}\left(\frac{m_\NS}{0.1\,\msun}\right),\quad
    f_\text{\tiny ISCO}\simeq4.4\,{\rm kHz}\left(\frac{\msun}{M_\BH}\right)\,,
    \label{eq:fRocheNSBH_gen}
\end{equation}
since the central BH is expected to be slowly spinning in a magnetically arrested disk state, $\chi_\BH \lesssim 0.2$~\cite{1977MNRAS.179..433B,Lowell:2023kyu,Jacquemin-Ide:2023aax, Gottlieb:2023cgm}.
A $0.1\,\msun$ NS, for both polytropic and nuclear matter EoSs, is thus disrupted near the most sensitive frequencies of ground-based detectors, nearly independently of $M_\BH$; the ISCO comes first only for $M_\BH\gtrsim67\,\msun\,(0.1\,\msun/m_\NS)$.
Quark stars are instead disrupted only at $\sim1.6\,$kHz, typically after reaching the ISCO~\cite{Baibhav:2026iny}.

The low compactness of subsolar-mass NSs also enhances finite-size effects. The tidal deformability $\Lambda=\tfrac{2}{3}k_2(R_\NS/m_\NS)^5$, with $k_2$ the quadrupolar tidal Love number~\cite{Hinderer:2007mb,Flanagan:2007ix},  and the spin-induced quadrupole $\mathcal{Q}=-\bar{\mathcal{Q}}\,\chi_\NS^2m_\NS^3$, with $\bar{\mathcal{Q}}\simeq2.5\,R_\NS/m_\NS$~\cite{Laarakkers:1997hb,Yagi:2013awa} and $\chi_\NS$ the NS spin~\cite{Metzger:2024ujc,Lo:2010bj,Alexander:2008wv}, reach values
\begin{equation}
\Lambda\simeq6\times10^{10}\left(\frac{m_\NS}{0.1\,\msun}\right)^{-20/3}\!,\;\;\bar{\mathcal{Q}}\simeq600\left(\frac{m_\NS}{0.1\,\msun}\right)^{-4/3}
    \label{eq:LambdaQbar}
\end{equation}
for the polytropic EoS, compared with $\Lambda\sim10^2 - 10^3$ and $\bar{\mathcal{Q}}\simeq4 - 8$ for canonical NSs~\cite{LIGOScientific:2017vwq,Poisson:1997ha}. Since BHs have $\Lambda=0$ and $\bar{\mathcal{Q}}=1$~\cite{Binnington:2009bb,Damour:2009va,Hui:2020xxx}, either imprint would exclude a primordial origin~\cite{Franciolini:2021xbq,Crescimbeni:2024cwh,Crescimbeni:2024qrq,DeLuca:2024uju,Russo:2025azm,Corman:2026lbt}, although both are suppressed by powers of $q$ in a NS-BH binary.

The disk environment further contributes with residual gas effects such as migration~\cite{Baibhav:2026iny}, as well as an initial eccentricity up to $e_{\text{\tiny NS-BH},i}\simeq0.1$ from in-disk assembly~\cite{Wu:2026hth}, and possible spin-orbit resonances~\cite{Miao:2025tqu}.

\vspace{0.1cm}
\noindent{{\bf{\em NS-BH waveform.}}} 
We model the GW signal from the collapsar NS-BH coalescence using the frequency-domain \texttt{TaylorF2} PN template in the stationary-phase approximation~\cite{Cutler:1994ys,Damour:2000zb,Buonanno:2009zt,Blanchet:2013haa}, augmenting the point-particle waveform with corrections that account for the environment, the finite size of the NS, its tidal disruption, and the initial conditions inherited from the collapsar disk. The resulting waveform model, \sig, can be written as:
\begin{align}
    \tilde h_\sig(f;\bm{\theta})
    ={}& \tilde h_\pp(f;\bm{\theta})\,
    \mathcal{S}_*(f;\bm{\theta})\,
    \mathcal{S}_\text{\tiny end}(f;\bm{\theta}) \nonumber\\
    &\times\exp\big[{\rm i}\,\delta\psi(f;\bm{\theta})\big]\,,
    \label{eq:perturbed_waveform}
\end{align}
where $\bm{\theta}$ collects the source and environmental parameters,
\begin{align}
    \label{eq:theta}
    \bm{\theta} = \big\{&\mathcal{M}_c,\,\eta,\,\chi_\BH,\,\chi_\NS,\,\Lambda,\,d_\text{\tiny L},\,\iota,\,t_c,\,\phi_c, \nonumber\\
    &a_{\text{\tiny NS-BH},i},\,e_{\text{\tiny NS-BH},i},\,f_\text{\tiny disk}\big\}\,.
\end{align}
The point-particle waveform reads
\begin{equation}\label{eq:pointparticle}
    \tilde h_\pp(f;\bm{\theta}) = C_\Omega(\bm{\theta})\,\mathcal{A}(f;\bm{\theta})\,\exp\big[{\rm i}\,\psi_\pp(f;\bm{\theta})\big]\,,
\end{equation}
with  amplitude $\mathcal{A}=\sqrt{5/24}\,\pi^{-2/3}\mathcal{M}_c^{5/6}d_\text{\tiny L}^{-1}f^{-7/6}$ in the Newtonian limit, where $d_\text{\tiny L}$ is the luminosity distance, $t_c$ and $\phi_c$ the time and phase at coalescence, $\mathcal{M}_c  \simeq M_\BH^{2/5} m_\NS^{3/5}$ the binary chirp mass, and $\psi_\pp$ the phase up to 3.5PN~\cite{Buonanno:2009zt,Blanchet:2013haa} depending on the masses and, through spin-orbit and spin-spin terms, on the aligned spins. We write it in terms of the PN velocity $v=(\pi Mf)^{1/3}$, the total mass $M=M_\BH+m_\NS$, and the symmetric mass ratio $\eta=M_\BH m_\NS/M^2$; its leading Newtonian term, $\psi_\text{\tiny N}=3/(128\,\eta\,v^5)$, sets the normalization of the phase corrections below. The detector response $C_\Omega$ depends on the inclination $\iota$ and the antenna patterns~\cite{Jaranowski:1994xd}; we take an optimally oriented, face-on binary, for which $C_\Omega=1$. 

The low-frequency window $\mathcal{S}_*=\Theta(f-f_{\text{\tiny NS-BH},i})$, with $\Theta$ the Heaviside function, ensures that the signal starts at the frequency of Eq.~\eqref{eq:fNSBHi}. At high frequencies, the window $\mathcal{S}_\text{\tiny end}(f;\bm{\theta})$  ends the signal: it is a tapered logistic function around $f_\text{\tiny Roche}$~\cite{DeLuca:2022xlz} if the NS is tidally disrupted, i.e. $f_\text{\tiny Roche}<f_\text{\tiny ISCO}$, with $f_\text{\tiny ISCO}$ the GW frequency at the prograde ISCO of the Kerr BH~\cite{Bardeen:1972fi}. Otherwise the NS plunges intact ($\mathcal{S}_\text{\tiny end} =1$), and beyond $f_\text{\tiny ISCO}$ the amplitude follows an $f^{-2/3}$ power law up to the ringdown frequency of the fundamental quasinormal mode of the remnant, and a Lorentzian above it~\cite{Flanagan:1997sx,Berti:2005ys,Ajith:2007kx,Pannarale:2015jka,Pannarale:2015jia}.

We consider four phase corrections, ordered from low to high frequencies:
\begin{equation}
    \label{eq:dpsi_total}
    \delta\psi = \delta\psi_\text{\tiny env} + \delta\psi_\text{\tiny ecc} + \delta\psi_\text{\tiny SIQM} + \delta\psi_\text{\tiny tidal}\,.
\end{equation}
Residual disk gas dephases the inspiral through Type~I migration~\cite{Kocsis:2011dr,Yunes:2011ws,Derdzinski:2018qzv,Zwick:2021dlg,Speri:2022upm,Baibhav:2026iny},
\begin{equation}
    \label{eq:dpsi_env}
    \delta\psi_\text{\tiny env}=-\frac{25\,C_\text{\tiny I}}{2496\,\pi}\,\frac{f_\text{\tiny disk}\,\dot M_\BH}{\alpha\,\mathbb{H}^4}\,v^{-8}\,\psi_\text{\tiny N}\,,
\end{equation}
with $C_\text{\tiny I}\simeq2.1$~\cite{2017MNRAS.471.4917J} and $\alpha \simeq 0.03$ the disk viscosity~\cite{Siegel:2018zxq}. Type~I migration is the disk analogue of dynamical friction, with rotation turning the wake into spiral density waves. 
The factor $f_\text{\tiny disk}$ rescales the density of the gas surrounding the NS at formation, with $f_\text{\tiny disk}=1$ corresponding to its configuration before fragmentation~\cite{Baibhav:2026iny}.  Fragmentation removes most of the gas from the outer regions and leaves the inner ones stable. In practice, $f_\text{\tiny disk}$ varies during the inspiral. Accretion onto the BH~\cite{Baibhav:2026iny} and outflows deplete the remaining gas~\cite{Metzger:2008av,Guo:2025sjb}, so $f_\text{\tiny disk}$ decreases with time.  
Given the poorly constrained post-fragmentation evolution of the disk, we parametrize it with a constant parameter $f_\text{\tiny disk}$ 
($f_\text{\tiny disk}=0$ in vacuum).
Equation~\eqref{eq:dpsi_env} applies while migration is subdominant to GW emission, $\dot f_\text{\tiny mig}<\dot f_\text{\tiny GW}$, i.e. for~\cite{Baibhav:2026iny}
\begin{equation}
    f\gtrsim26\,{\rm Hz}\left(\frac{f_\text{\tiny disk}}{10^{-3}}\,\frac{\dot M_\BH}{0.1\,\msun\,{\rm s}^{-1}}\right)^{3/8}\left(\frac{0.3}{\mathbb{H}}\right)^{3/2}\frac{5\,\msun}{M_\BH}\,.
    \label{eq:fmig}
\end{equation}
This holds across the \textsc{GstLAL} band for $f_\text{\tiny disk}\leq10^{-3}$ and $M_\BH\gtrsim3\,\msun$, so we treat $f_\text{\tiny disk}=10^{-3}$ as a limiting case.
In principle, Bondi accretion onto the NS gives rise to a correction at the same $-4\,{\rm PN}$ order as migration~\cite{Kocsis:2011dr,Barausse:2014tra,CanevaSantoro:2023aol,Baibhav:2026iny}, but its magnitude depends not only on $f_\text{\tiny disk}$ but also on the accretion efficiency, and on whether the accreted gas carries net orbital angular momentum about the binary's center of mass. Given this uncertainty and its degeneracy with migration, we neglect this contribution; Eq.~\eqref{eq:dpsi_env} therefore provides a conservative estimate of the environmental dephasing.

A residual eccentricity $e_{\text{\tiny NS-BH},i}$ at $f_{\text{\tiny NS-BH},i}$ adds, to leading order in $e\ll1$, the $-19/6$\,PN correction~\cite{Yunes:2009yz,Favata:2013rwa,Moore:2016qxz,Favata:2021vhw}
\begin{equation}
    \label{eq:dpsi_ecc}
    \delta\psi_\text{\tiny ecc} = -\frac{2355}{1462}\,e_{\text{\tiny NS-BH},i}^2\left(\frac{v_i}{v}\right)^{19/3}\psi_\text{\tiny N}\,,
\end{equation}
with $v_i = (\pi M f_{\text{\tiny NS-BH},i})^{1/3}$, which accumulates over the long in-band inspiral even for small eccentricities~\cite{Breivik:2016ddj,Nishizawa:2016eza,Zevin:2021rtf,Romero-Shaw:2022xko}. Close encounters among the fragments, however, can keep many NSs highly eccentric in band~\cite{ Baibhav:2026zuy}, requiring fully eccentric waveforms with higher orbital harmonics and amplitude modulations~\cite{Huerta:2014eca,Klein:2018ybm,Moore:2018kvz}.

The quadrupole-monopole interaction between the BH and the spin-induced NS quadrupole enters at 2PN order, with the deviation from the Kerr value~\cite{Poisson:1997ha,Krishnendu:2017shb,Krishnendu:2018nqa}
\begin{equation}
    \label{eq:dpsi_quadrupole}
    \delta\psi_\text{\tiny SIQM}
    = -\frac{75}{64}\,q\,
    \frac{(\bar{\mathcal{Q}}-1)\,\chi_\NS^{2}}{v}\,.
\end{equation}
Finally, the tidal deformability enters the waveform starting at 5PN order~\cite{Flanagan:2007ix,Hinderer:2009ca,Vines:2011ud,Wade:2014vqa,LIGOScientific:2017vwq},
\begin{align}
    \label{eq:psi_tidal}
    \delta\psi_\text{\tiny tidal}
    &=  -24\, \frac{12+q}{(1+q)^5}\, q^4 \Lambda\,v^{10}  \psi_\text{\tiny N}\,.
\end{align}
Our waveform also includes the 6PN tidal term~\cite{Vines:2011ud} (see Refs.~\cite{Pani:2025qxs, Apostolidis:2026qsg, Valletta:2026loz} for higher-order PN corrections to the tidal phase).

\begin{figure*}[!t]
\includegraphics[width=\linewidth]{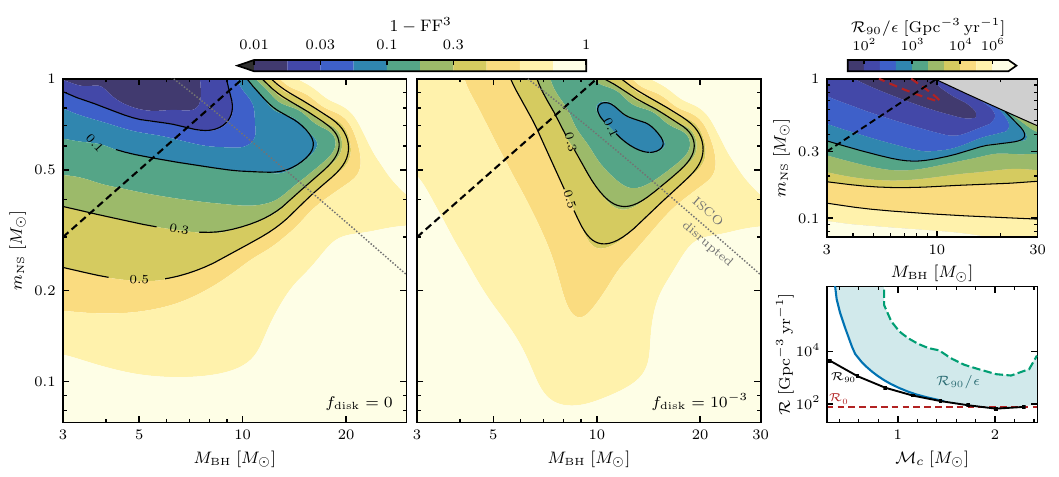}
\caption{
Left and middle columns: 
fraction of collapsar NS-BH events missed by the \textsc{GstLAL} O4 subsolar search, $1-{\rm FF}^3$ [Eq.~\eqref{eq:FF}], for $f_\text{\tiny disk}=0$ and $10^{-3}$. The black dashed line marks the LVK template-bank boundary at $q=0.1$, and the gray dotted line marks the curve $f_\text{\tiny Roche}=f_\text{\tiny ISCO}$ [Eq.~\eqref{eq:fRocheNSBH_gen}], above which the NS reaches the ISCO before being disrupted. For $m_\NS \lesssim 0.073\,\msun$, the NS is disrupted before the GW frequency reaches $f_\text{\tiny min} = 45\,$Hz.
Right column: upper limit on the collapsar NS-BH merger rate, $\mathcal{R}_{90}/\epsilon$ [Eq.~\eqref{eq:eps}], from the \textsc{GstLAL} O4a+b limits $\mathcal{R}_{90}$~\cite{LIGOScientific:2026kef}, with $\epsilon$ relative to template waveforms of the same chirp mass ($f_\text{\tiny disk}=0$). The top panel shows this limit in the $(M_\BH,m_\NS)$ plane. The red dashed contour marks $\mathcal{R}_{90}/\epsilon = \mathcal{R}_0$, where $\mathcal{R}_0$ is the LGRB rate, and the gray contour marks the boundary $\mathcal{M}_c = 2.5\,\msun$ of the last LVK chirp-mass bin, beyond which no limit is available. The bottom panel shows $\mathcal{R}_{90}/\epsilon$ versus $\mathcal{M}_c$ for $M_\BH=3\,\msun$ (blue) and $30\,\msun$ (green, dashed), with $m_\NS$ from $\simeq0.07$ to $1\,\msun$; the shaded band covers intermediate values of $M_\BH$. Black squares mark $\mathcal{R}_{90}$ at the bin centers, and the red dashed line indicates $\mathcal{R}_0$.
} 
\label{fig:ssm_loss}
\end{figure*}

\vspace{0.1cm}
\noindent{{\bf{\em Fitting factor and lost events.}}}
Matched-filtering searches correlate the data with a template bank $h_\text{\tiny T}(\bm{\lambda})$ that spans the source parameters $\bm{\lambda}$. If the template bank does not contain an accurate representation of a signal, the recovered signal-to-noise ratio (SNR) is reduced. This reduction is quantified by the fitting factor~\cite{Apostolatos:1995pj} 
\begin{equation}
    \label{eq:FF}
    {\rm FF}(\bm{\theta})=\max_{\bm{\lambda},\,t_c,\,\phi_c}\frac{\langle h_\sig(\bm{\theta})|h_\text{\tiny T}(\bm{\lambda})\rangle}{\sqrt{\langle h_\sig|h_\sig\rangle\,\langle h_\text{\tiny T}|h_\text{\tiny T}\rangle}}\,,
\end{equation}
where $\langle\cdot|\cdot\rangle$ denotes the inner product weighted by the O4 noise spectrum over the search band~\cite{TheLIGOScientific:2014jea,Aasi:2013wya}. A matched-filtering template that exactly matches the true signal yields the optimal SNR $\rho_\sig=\langle h_\sig|h_\sig\rangle^{1/2}$; the maximum recovered SNR is ${\rm FF}\times \rho_\sig$. 
We maximize over $\phi_c$ analytically, over $t_c$ with an inverse fast Fourier transform~\cite{Allen:2005fk}, and over $\bm{\lambda}$ numerically.

We adopt the \textsc{GstLAL} configuration of the O4 subsolar search, which filters the data above $f_\text{\tiny min}=45\,$Hz with aligned-spin waveforms~\cite{LIGOScientific:2026kef}. The template bank, the waveform approximants, and the other pipelines are described in the Supplemental Material. We consider non-spinning, circular signals with the tidal phase. To quantify the effect of residual gas in the collapsar model, we include $\delta\psi_\text{\tiny env}$ with $f_\text{\tiny disk}=10^{-3}$, $\mathbb{H}=0.3$, and $\dot M_\BH\simeq0.1\,\msun\,{\rm s}^{-1}$.

A search reports a detection when the recovered SNR exceeds a fixed threshold. Because the SNR decreases inversely with the luminosity distance, a source with recovered SNR ${\rm FF}\times\rho_\sig$ is detectable out to a horizon distance proportional to ${\rm FF}\times\rho_\sig$. For sources distributed uniformly in the local volume, the number of detected events grows as the cube of this distance. Compared with an ideal search that filters each signal with its exact waveform (${\rm FF}=1$) over the same band, a search with imperfect templates therefore detects only a fraction ${\rm FF}^3$ of the events, and misses a fraction $1-{\rm FF}^3$.

\vspace{0.1cm}
\noindent{{\bf{\em Event losses in the O4 subsolar search.}}}
The event loss results are summarized in Fig.~\ref{fig:ssm_loss}.
Within the applicability region of the template bank, the missed fraction is only a few percent for heavy NSs orbiting light BHs (left column). It increases as the NS mass decreases, reaching tens of percent at $m_\NS\sim0.3\,\msun$ and $\gtrsim90\%$ at $m_\NS\sim0.1\,\msun$. The clumps described by Eq.~\eqref{eq:clumpmass} have mass ratios $q\sim10^{-3}-10^{-1}$, potentially below the template-bank boundary. Consequently, the optimal $q=0.1$ template accumulates a 1PN-order phase difference relative to the signal over $N_\text{\tiny cyc}\propto\mathcal{M}_c^{-5/3}f_\text{\tiny min}^{-5/3}$ cycles; this number increases as $m_\NS$ decreases. Also, light NSs are disrupted at $f_\text{\tiny Roche}\propto m_\NS$, only $\mathcal{O}(10)\,$Hz above $f_\text{\tiny min}$ for all values of $M_\BH$ considered here, whereas the templates extend through merger and therefore contain high-frequency SNR that is absent from the signal.

For $q<0.1$, the detection efficiency remains high when $m_\NS\gtrsim0.3\,\msun$. We find that optimal templates have $q=0.1$ and prefer near-extremal, anti-aligned primary spins. An anti-aligned spin moves the ISCO outward, lowering the frequency at which the template inspiral ends toward the disruption frequency of the signal, and accelerates the late-inspiral frequency evolution; both partially compensate for the smaller mass ratio of the signal and its termination at tidal disruption. For $M_\BH>10\,\msun$, templates with the same chirp mass continue to provide high fitting factors for heavy NSs.

Relative to the GW-driven frequency evolution, the migration-induced phase correction scales as $(M_\BH f)^{-8/3}$ [Eq.~\eqref{eq:dpsi_env}]. Its contribution is therefore largest at low frequencies and for light BHs, where the vacuum fitting factors are highest (middle column). In this region, an in-band dephasing of $\sim10 - 10^3\,$rad increases the missed fraction from a few percent to more than $50\%$. By contrast, the missed fraction remains $\sim10\%$ for heavy NSs orbiting BHs with $M_\BH\simeq (10 - 20)\,\msun$. Like in the vacuum case, the template models fail to capture the regions of heavy BH or light NS masses, so a large fraction of events is missed (see Supplemental Material for details on the template bank and on the contribution of each effect).

In the absence of significant candidates, the O4 subsolar search constrains the merger rate at $90\%$ confidence over chirp masses spanning $(0.16 - 2.44)\,\msun$, with $\mathcal{R}_{90}=2.3/\langle VT\rangle$ and $\langle VT\rangle$ the surveyed volume-time of the search~\cite{Brady:2004gt}.
We interpolate the \textsc{GstLAL} O4a+O4b limits in $\log\mathcal{R}_{90}$ as a function of $\log\mathcal{M}_c$. 
To apply these limits to collapsar events, we need the sensitive volume-time of the search for our signal, whereas the $\langle VT \rangle$ reported by the LVK~\cite{LIGOScientific:2026kef} is estimated from an injection campaign based on binary BH signals. As a proxy for the LVK injection model, we therefore adopt the point-particle waveform $h_0$, whose Fourier transform is given in Eq.~\eqref{eq:pointparticle}. Besides the template mismatch, which retains a fraction ${\rm FF}^3$ of the events, the collapsar signal is intrinsically quieter than a BH binary at the same masses and distance, because it ends at tidal disruption and therefore lacks the late inspiral and merger, which contribute a good portion of the SNR. We quantify this by the ratio of the optimal SNRs over the search band,
\begin{equation}
    B\equiv\frac{\rho_\sig}{\rho_\pp}\,,
    \label{eq:B}
\end{equation}
where $\rho_\pp = \langle h_\pp|h_\pp\rangle^{1/2}$ is the optimal SNR of the point-particle waveform $h_\pp$ with the same component masses.  Even with ${\rm FF}=1$, 
a collapsar source would therefore be detected only out to a fraction $B$ of the horizon distance of the LVK injected waveforms. The recovered SNR of a collapsar source is then ${\rm FF}\times B\times\rho_\pp$, and by the same argument the search detects, relative to the injected BH binaries, a fraction
\begin{equation}
    \epsilon=\left({\rm FF}\times B\right)^3
    \label{eq:eps}
\end{equation}
of the events. The sensitive volume-time for collapsar signals is thus $\epsilon\langle VT\rangle$, and the upper limit on the collapsar channel becomes $\mathcal{R}_{90}/\epsilon$ (right column of Fig.~\ref{fig:ssm_loss}).
Relative to the published limit, the bound is weaker by orders of magnitude for light NSs, for which $\epsilon\sim10^{-2}$, and it is comparable for heavy NSs orbiting BHs with $M_\BH\simeq (5 - 10)\,\msun$. It equals the LGRB rate $\mathcal{R}_0\simeq80\,{\rm Gpc^{-3}\,yr^{-1}}$~\cite{Ghirlanda:2022edk}, corresponding to one NS-BH merger per collapsar, only near $\mathcal{M}_c\simeq2\,\msun$. These weaker limits also apply to nuclear-matter EoSs, whereas much stronger limits are expected for quark-matter stars ($B \simeq 1$), which reach higher frequencies before disruption.

\emph{Caveats.}
The rescaling in Eq.~\eqref{eq:eps} assumes that the ratio of sensitive volume-times of collapsar signals and of the LVK injections in each chirp-mass bin, $\langle VT\rangle_\sig/\langle VT\rangle_0$, equals $\epsilon$. The injections, however, span a range of component masses and mass ratios $q\geq0.05$ within each bin, whereas our sources have $q\sim10^{-2}$, outside the injected range. Replacing the injection-averaged sensitivity with that of our reference waveform changes $\mathcal{R}_{90}/\epsilon$ by a factor $\lesssim 1.5$; the curves in Fig.~\ref{fig:ssm_loss} are therefore order-of-magnitude estimates based on the published limits, not bounds derived from the search itself.
In addition, actual searches use a discrete bank~\cite{Owen:1998dk}, ranking statistics~\cite{Cannon:2015gha,Nitz:2017svb,Tsukada:2023edh}, signal-consistency tests~\cite{Allen:2004gu,Messick:2016aqy,DalCanton:2014hxh}, and coincidence requirements~\cite{Usman:2015kfa,Davies:2020tsx}, which penalize mismatched and short, narrowband signals beyond their SNR loss. 
Our losses are therefore idealized estimates of the sensitive-volume reduction, and are approximate bounds on the actual losses, that should be estimated through injections into the search pipelines. For instance, for a signal that does not match the template, the $\chi^2$ statistic acquires a non-centrality parameter scaling as $(1-{\rm FF}^2)\, \rho_\sig^2$~\cite{Allen:2004gu}, so that loud events, which the fitting factor alone would deem recoverable, are further down-weighted.

 The waveform model adopted here is subject to additional theoretical uncertainties. For mass ratios $q\sim10^{-2}$ the signal spans $10^4 - 10^5$ cycles in band, where PN truncation errors and self-force corrections can accumulate, and no numerical relativity simulations exist in this regime to calibrate the tidal disruption, the merger, or the environmental terms. We also assume that the signal ends at Roche-lobe overflow. For light NSs, mass transfer onto the BH may instead be stable~\cite{Clark:1977,PortegiesZwart:1998eca}, so that the NS is stripped over many orbits and keeps emitting near $f_\text{\tiny Roche}$, with a frequency evolution that need not be monotonic, as found for white dwarfs stripped by massive BHs~\cite{Zalamea:2010mv}. Such a phase would extend the signal beyond the termination assumed here, which accounts for a large part of the losses, and would carry its own signature. Our results should therefore be read as order-of-magnitude estimates of the losses; reliable searches for these sources will require dedicated waveform models that incorporate these effects and are validated against numerical simulations.

\vspace{0.1cm}
\noindent{{\bf{\em Conclusions.}}} 
A confirmed detection of a subsolar component in a binary coalescence would revolutionize our understanding of compact object formation. Besides PBHs and exotic compact objects, the strongest astrophysical alternative is provided by collapsars, where the outer, neutrino-cooled accretion disk around the newborn BH fragments into subsolar NSs that can later merge with the central BH.

In this work we have built a waveform model, \sig, for these NS-BH binaries, augmenting the \texttt{TaylorF2} template with the main features inherited from the collapsar environment: a tidal-disruption terminator, the enhanced spin-induced quadrupole and tidal deformability of subsolar NSs, residual eccentricity, and the $-4$PN gas-driven dephasing. These features make their signals markedly different from the templates of current subsolar searches, which describe quasicircular, nearly equal-mass $q \geq 0.1$ binaries without environmental corrections, with tides included only for light binaries.

No statistically significant candidates were identified so far within the LVK fourth observing run, constraining their merger rate, but such upper limits should be taken with a grain of salt when compared with NS-BHs from the collapsar channel, which may be missed by the searches.
To quantify this, we have computed the fitting
factor of our waveform against the LVK O4b template family, estimating the fraction of collapsar events that current searches would miss, and provided approximate bounds on the merger rates.  We find that the search misses up to $30\%$ of the events for $q>0.1$, and more than $90\%$ at $m_\NS\sim0.1\,\msun$, because of the small mass ratios and the early tidal disruption; residual gas increases these losses further around light BHs. Accordingly, the upper limits on the merger rate weaken by orders of magnitude for light NSs, and constrain the collapsar channel at the level of the LGRB rate only near $\mathcal{M}_c\simeq2\,\msun$. The O4 searches reported several marginal subsolar candidates. If these were collapsar mergers, the recovered SNR would underestimate the optimal SNR by the fitting factor ${\rm FF}$: for light NSs, with ${\rm FF}\sim0.3 - 0.5$, the optimal SNR would be a factor of $2-3$ larger, so a candidate near the detection threshold could correspond to a louder signal. Refiltering such candidates with waveforms that include the effects discussed here could therefore increase their significance. Moreover, decreasing the low-frequency cutoff of the templates would also increase the number of events: relative to a search starting at $20\,$Hz, the $45\,$Hz cutoff of \textsc{GstLAL} loses $\sim80\%$ of the events at $m_\NS\simeq0.1\,\msun$, compared with $\sim30\%$ for the $30\,$Hz cutoff of \textsc{PyCBC} (see Supplemental Material).
 Should a candidate be found, its association with the accompanying LGRB or broad-lined supernova, or with subsolar BNS mergers from the same collapsar~\cite{Baibhav:2026iny}, would provide a smoking-gun test of its origin.

Our work can be extended in several directions. A dedicated search for these sources in current and future GW data requires a template bank covering the parameter space of the model, which in turn demands a careful treatment of their features dictated by the collapsar environment. Along these lines, it will be important to go beyond the leading-order description adopted here, including the impact of potentially large eccentricities arising from the complex $N$-body dynamics of the fragments, higher harmonics excited by the intermediate mass ratios (particularly when tidal disruption is ineffective), environmental corrections beyond leading order from the residual gas around the central BH, and possible spin-orbit resonances of the NS. We plan to return to these points in future work.

\vspace{0.1cm}
\noindent{{\bf{\em Acknowledgments.}}}
We thank Mark Cheung, Neil Cornish, and Digvijay Wadekar for useful discussions.
V.B. acknowledges support from the NASA Hubble Fellowship grant HST-HF2-51548.001-A awarded by the Space Telescope Science Institute, which
is operated by the Association of Universities for Research
in Astronomy, Inc., for NASA, under contract NAS5-26555.
E.B, V.D.L. and L.D.G. are supported by NSF Grants No.~AST-2606672, No.~PHY-2513337, No.~PHY-090003, and No.~PHY-20043, by the Simons Foundation [MPS-SIP-00001698, E.B.], by the Simons Foundation International [SFI-MPS-BH-00012593-02], and by Italian Ministry of Foreign Affairs and International Cooperation Grant No.~PGR01167.
This work used Delta at the National Center for Supercomputing Applications (NCSA) through allocation PHY260065 from the Advanced Cyberinfrastructure Coordination Ecosystem: Services \& Support (ACCESS) program, which is supported by U.S. National Science Foundation grants 
\#2138259, \#2138286, \#2138307, \#2137603, and \#2138296.
This work was carried out at the Advanced Research Computing at Hopkins (ARCH) core facility (\url{https://www.arch.jhu.edu/}), which is supported by the NSF Grant No.~OAC-1920103.
B.D.M. acknowledges support from the National Science Foundation (grant AST-2406637), NASA (grant 80NSSC26K0299), and the Simons Foundation
(grant 727700).

\bibliography{draft}

@article{Baibhav:2026zuy,
    author = "Baibhav, Vishal and Metzger, Brian D. and Hui, Lam",
    title = "{A Collapsar-disk Origin for GW190814}",
    eprint = "2606.23786",
    archivePrefix = "arXiv",
    primaryClass = "astro-ph.HE",
    doi = "10.3847/2041-8213/aeb53c",
    journal = "Astrophys. J. Lett.",
    volume = "1011",
    number = "2",
    pages = "L49",
    year = "2026"
}

@article{Phukon:2021cus,
    author = "Phukon, Khun Sang and Baltus, Gregory and Caudill, Sarah and Clesse, Sebastien and Depasse, Antoine and Fays, Maxime and Fong, Heather and Kapadia, Shasvath J. and Magee, Ryan and Tanasijczuk, Andres Jorge",
    title = "{Search for subsolar primordial black holes in low mass-ratio binaries with LIGO-Virgo O2 data and implications for the primordial black hole dark matter fraction}",
    eprint = "2105.11449",
    archivePrefix = "arXiv",
    primaryClass = "astro-ph.CO",
    reportNumber = "LIGO-P2100172",
    doi = "10.1103/cpjc-qvwl",
    journal = "Phys. Rev. D",
    volume = "114",
    number = "6",
    pages = "063529",
    year = "2026"
}

@article{Messick:2016aqy,
    author = "Messick, Cody and others",
    title = "{Analysis Framework for the Prompt Discovery of Compact Binary Mergers in Gravitational-wave Data}",
    eprint = "1604.04324",
    archivePrefix = "arXiv",
    primaryClass = "astro-ph.IM",
    doi = "10.1103/PhysRevD.95.042001",
    journal = "Phys. Rev. D",
    volume = "95",
    number = "4",
    pages = "042001",
    year = "2017"
}

@article{Guo:2025sjb,
    author = "Guo, Minghao and Stone, James M. and Quataert, Eliot and Springel, Volker",
    title = "{Cyclic Zoom: Multiscale GRMHD Modeling of Black Hole Accretion and Feedback}",
    eprint = "2504.16802",
    archivePrefix = "arXiv",
    primaryClass = "astro-ph.HE",
    month = "4",
    year = "2025"
}

@article{Metzger:2008av,
    author = "Metzger, B. D. and Piro, A. L. and Quataert, E.",
    title = "{Time-Dependent Models of Accretion Disks Formed from Compact Object Mergers}",
    eprint = "0805.4415",
    archivePrefix = "arXiv",
    primaryClass = "astro-ph",
    doi = "10.1111/j.1365-2966.2008.13789.x",
    journal = "Mon. Not. Roy. Astron. Soc.",
    volume = "390",
    pages = "781",
    year = "2008"
}

@article{Brady:2004gt,
    author = "Brady, Patrick R. and Creighton, Jolien D. E. and Wiseman, Alan G.",
    title = "{Upper limits on gravitational-wave signals based on loudest events}",
    eprint = "gr-qc/0405044",
    archivePrefix = "arXiv",
    doi = "10.1088/0264-9381/21/20/020",
    journal = "Class. Quant. Grav.",
    volume = "21",
    pages = "S1775--S1782",
    year = "2004"
}

@ARTICLE{1991ApJ...369..422C,
       author = {{Colpi}, Monica and {Shapiro}, Stuart L. and {Teukolsky}, Saul A.},
        title = "{Explosion of a Rotating Neutron Star near the Minimum Mass}",
      journal = {\apj},
         year = 1991,
        month = mar,
       volume = {369},
        pages = {422},
          doi = {10.1086/169771},
       adsurl = {https://ui.adsabs.harvard.edu/abs/1991ApJ...369..422C}
}

@article{Moore:2018kvz,
    author = "Moore, Blake and Robson, Travis and Loutrel, Nicholas and Yunes, Nicolas",
    title = "{Towards a Fourier domain waveform for non-spinning binaries with arbitrary eccentricity}",
    eprint = "1807.07163",
    archivePrefix = "arXiv",
    primaryClass = "gr-qc",
    doi = "10.1088/1361-6382/aaea00",
    journal = "Class. Quant. Grav.",
    volume = "35",
    number = "23",
    pages = "235006",
    year = "2018"
}

@article{Klein:2018ybm,
    author = "Klein, Antoine and Boetzel, Yannick and Gopakumar, Achamveedu and Jetzer, Philippe and de Vittori, Lorenzo",
    title = "{Fourier domain gravitational waveforms for precessing eccentric binaries}",
    eprint = "1801.08542",
    archivePrefix = "arXiv",
    primaryClass = "gr-qc",
    doi = "10.1103/PhysRevD.98.104043",
    journal = "Phys. Rev. D",
    volume = "98",
    number = "10",
    pages = "104043",
    year = "2018"
}

@article{Ackley:2026zlf,
    author = "Ackley, Kendall and others",
    title = "{ENGRAVE follow-up of a type IIb supernova spatially coincident with the sub-threshold gravitational wave trigger S250818k}",
    eprint = "2605.02639",
    archivePrefix = "arXiv",
    primaryClass = "astro-ph.HE",
    doi = "10.1051/0004-6361/202660729",
    journal = "Astron. Astrophys.",
    volume = "714",
    pages = "A284",
    year = "2026"
}

@article{Golomb:2024mmt,
    author = "Golomb, Jacob and Legred, Isaac and Chatziioannou, Katerina and Abac, Adrian and Dietrich, Tim",
    title = "{Using equation of state constraints to classify low-mass compact binary mergers}",
    eprint = "2403.07697",
    archivePrefix = "arXiv",
    primaryClass = "astro-ph.HE",
    doi = "10.1103/PhysRevD.110.063014",
    journal = "Phys. Rev. D",
    volume = "110",
    number = "6",
    pages = "063014",
    year = "2024"
}

@article{Apostolatos:1995pj,
    author = "Apostolatos, T. A.",
    title = "{Search templates for gravitational waves from precessing, inspiraling binaries}",
    doi = "10.1103/PhysRevD.52.605",
    journal = "Phys. Rev. D",
    volume = "52",
    pages = "605--620",
    year = "1995"
}

@article{RamisVidal:2026ycb,
    author = "Ramis Vidal, Felip A. and Abac, Adrian and Colleoni, Marta and Dietrich, Tim and Mourier, Pierre and Gonzalez, Alejandra and Markin, Ivan and Puecher, Anna",
    title = "{Fast gravitational waveform models for quasicircular coalescences of neutron star-black hole binaries}",
    eprint = "2606.04810",
    archivePrefix = "arXiv",
    primaryClass = "gr-qc",
    doi = "10.1103/b3c9-s8lh",
    journal = "Phys. Rev. D",
    volume = "114",
    number = "6",
    pages = "064044",
    year = "2026"
}

@article{Markin:2026eyc,
    author = "Markin, Ivan and Bulla, Mattia and Dietrich, Tim",
    title = "{Numerical simulations of black hole-neutron star mergers with equal and near-equal mass ratios}",
    eprint = "2601.19405",
    archivePrefix = "arXiv",
    primaryClass = "astro-ph.HE",
    doi = "10.1103/bjs5-ytrp",
    journal = "Phys. Rev. D",
    volume = "114",
    number = "2",
    pages = "024008",
    year = "2026"
}

@article{Cheung:2025grp,
    author = "Cheung, Mark Ho-Yeuk and Wadekar, Digvijay and Mehta, Ajit Kumar and Islam, Tousif and Roulet, Javier and Berti, Emanuele and Venumadhav, Tejaswi and Zackay, Barak and Zaldarriaga, Matias",
    title = "{Searching for intermediate mass ratio binary black hole mergers in the third observing run of LIGO-Virgo-KAGRA}",
    eprint = "2507.01083",
    archivePrefix = "arXiv",
    primaryClass = "gr-qc",
    doi = "10.1103/l3ql-mv6v",
    journal = "Phys. Rev. D",
    volume = "113",
    number = "2",
    pages = "023003",
    year = "2026"
}

@article{Kacanja:2026nwi,
    author = "Kacanja, Keisi and Soni, Kanchan and Nitz, Alexander H.",
    title = "{Search for Sub-Solar Mass Binaries in the First and Second Part of LIGO's Fourth Observing Run O4a-O4b}",
    eprint = "2610.19329",
    archivePrefix = "arXiv",
    primaryClass = "astro-ph.HE",
    month = "10",
    year = "2026"
}

@article{Baibhav:2026iny,
    author = "Baibhav, Vishal and De Luca, Valerio and Del Grosso, Loris and Berti, Emanuele and Metzger, Brian D. and Hui, Lam",
    title = "{Gravitational wave signatures of subsolar neutron star formation in collapsar disks}",
    eprint = "2609.25225",
    archivePrefix = "arXiv",
    primaryClass = "gr-qc",
    month = "9",
    year = "2026"
}

@article{Virgo:2025aai,
    author = "Abac, A. G. and others",
    collaboration = "Virgo, KAGRA, LIGO Scientific",
    title = "{Upper limits on the isotropic gravitational-wave background from the first part of LIGO, Virgo, and KAGRA{\textquoteright}s fourth observing run}",
    eprint = "2508.20721",
    archivePrefix = "arXiv",
    primaryClass = "gr-qc",
    reportNumber = "LIGO-P2500349",
    doi = "10.1103/wq57-sjt2",
    journal = "Phys. Rev. D",
    volume = "114",
    number = "4",
    pages = "042001",
    year = "2026"
}

@article{LIGOScientific:2026kef,
    author = "Abac, A. G. and others",
    collaboration = "LIGO Scientific, VIRGO, KAGRA",
    title = "{Search for Compact Binary Coalescences with Sub-solar Mass Components in Data from the Second Part of LIGO-Virgo-KAGRA's Fourth Observing Run}",
    eprint = "2610.20276",
    archivePrefix = "arXiv",
    primaryClass = "gr-qc",
    month = "10",
    year = "2026"
}

@article{DeLuca:2026eis,
    author = "De Luca, Valerio and Del Grosso, Loris and Berti, Emanuele",
    title = "{Stochastic gravitational-wave background from subsolar neutron stars in collapsar disks}",
    eprint = "2609.36005",
    archivePrefix = "arXiv",
    primaryClass = "gr-qc",
    month = "9",
    year = "2026"
}

@article{Corman:2026lbt,
    author = "Corman, Maxence and East, William E. and Read, Jocelyn S.",
    title = "{Binary neutron star mergers with a subsolar mass star}",
    eprint = "2603.25102",
    archivePrefix = "arXiv",
    primaryClass = "astro-ph.HE",
    doi = "10.1103/j5cs-3d6r",
    journal = "Phys. Rev. D",
    volume = "114",
    number = "6",
    pages = "063040",
    year = "2026"
}

@article{Miao:2025tqu,
    author = "Miao, Zhiqiang and Yang, Huan",
    title = "{Probing intrinsic ellipticity in compact-star binaries}",
    eprint = "2511.08895",
    archivePrefix = "arXiv",
    primaryClass = "astro-ph.HE",
    doi = "10.1103/mz9l-z23x",
    journal = "Phys. Rev. D",
    volume = "114",
    number = "6",
    pages = "064080",
    year = "2026"
}

@article{Karim:2026bva,
    author = "Karim, Waly M. Z. and Burrows, Adam and Vartanyan, David and Fisher, Anthony and Burry, James",
    title = "{The Mass Function of Neutron Stars from Core-Collapse Supernova Simulations}",
    eprint = "2608.18198",
    archivePrefix = "arXiv",
    primaryClass = "astro-ph.HE",
    month = "8",
    year = "2026"
}

@article{Ghirlanda:2022edk,
    author = "Ghirlanda, G. and Salvaterra, R.",
    title = "{The Cosmic History of Long Gamma Ray Bursts}",
    eprint = "2206.06390",
    archivePrefix = "arXiv",
    primaryClass = "astro-ph.HE",
    month = "6",
    year = "2022"
}

@article{LIGOScientific:2025slb,
    author = "Abac, A. G. and others",
    collaboration = "LIGO Scientific, VIRGO, KAGRA",
    title = "{GWTC-4.0: Updating the Gravitational-Wave Transient Catalog with Observations from the First Part of the Fourth LIGO-Virgo-KAGRA Observing Run}",
    eprint = "2508.18082",
    archivePrefix = "arXiv",
    primaryClass = "gr-qc",
    reportNumber = "LIGO-P2400386",
    doi = "10.3847/2041-8213/ae2c74",
    journal = "Astrophys. J. Lett.",
    volume = "1004",
    number = "2",
    pages = "L22",
    year = "2026"
}

@article{Kacanja:2026byy,
    author = {Kacanja, Keisi and Soni, Kanchan and Aky{\"u}z, Aleyna and Nitz, Alexander H.},
    title = "{Search for subsolar mass binaries in the first part of LIGO{\textquoteright}s fourth observing run}",
    eprint = "2602.12115",
    archivePrefix = "arXiv",
    primaryClass = "astro-ph.HE",
    doi = "10.1103/ztpt-pwl5",
    journal = "Phys. Rev. D",
    volume = "114",
    number = "4",
    pages = "043030",
    year = "2026"
}

@article{Yunes:2009yz,
    author = "Yunes, Nicolas and Arun, K. G. and Berti, Emanuele and Will, Clifford M.",
    title = "{Post-Circular Expansion of Eccentric Binary Inspirals: Fourier-Domain Waveforms in the Stationary Phase Approximation}",
    eprint = "0906.0313",
    archivePrefix = "arXiv",
    primaryClass = "gr-qc",
    doi = "10.1103/PhysRevD.80.084001",
    journal = "Phys. Rev. D",
    volume = "80",
    number = "8",
    pages = "084001",
    year = "2009",
    note = "[Erratum: Phys.Rev.D 89, 109901 (2014)]"
}

@article{Huerta:2014eca,
    author = "Huerta, E. A. and Kumar, Prayush and McWilliams, Sean T. and O'Shaughnessy, Richard and Yunes, Nicol{\'a}s",
    title = "{Accurate and efficient waveforms for compact binaries on eccentric orbits}",
    eprint = "1408.3406",
    archivePrefix = "arXiv",
    primaryClass = "gr-qc",
    doi = "10.1103/PhysRevD.90.084016",
    journal = "Phys. Rev. D",
    volume = "90",
    number = "8",
    pages = "084016",
    year = "2014"
}

@article{Moore:2016qxz,
    author = "Moore, Blake and Favata, Marc and Arun, K. G. and Mishra, Chandra Kant",
    title = "{Gravitational-wave phasing for low-eccentricity inspiralling compact binaries to 3PN order}",
    eprint = "1605.00304",
    archivePrefix = "arXiv",
    primaryClass = "gr-qc",
    reportNumber = "LIGO-DCC-P1500268",
    doi = "10.1103/PhysRevD.93.124061",
    journal = "Phys. Rev. D",
    volume = "93",
    number = "12",
    pages = "124061",
    year = "2016"
}

@article{Liu:2026hmk,
    author = "Liu, Zhengyan and others",
    title = "{WFST Follow-up of S251112cm: Searching for an Optical Counterpart to a Subsolar-mass Compact-binary Merger Candidate}",
    eprint = "2609.18523",
    archivePrefix = "arXiv",
    primaryClass = "astro-ph.HE",
    doi = "10.3847/2041-8213/aeb6bf",
    journal = "Astrophys. J. Lett.",
    volume = "1011",
    number = "2",
    pages = "L50",
    year = "2026"
}

@article{LIGOScientific:2026wfs,
    collaboration = "LIGO Scientific, VIRGO, KAGRA",
    title = "{GWTC-5.0: Observations from the Second Part of the Fourth LIGO-Virgo-KAGRA Observing Run and Updates to the Gravitational-Wave Transient Catalog}",
    eprint = "2605.27225",
    archivePrefix = "arXiv",
    primaryClass = "gr-qc",
    reportNumber = "LIGO-P2600152",
    month = "5",
    year = "2026"
}

@article{Lowell:2023kyu,
    author = "Lowell, Beverly and Jacquemin-Ide, Jonatan and Tchekhovskoy, Alexander and Duncan, Alex",
    title = "{Rapid Black Hole Spin-down by Thick Magnetically Arrested Disks}",
    eprint = "2302.01351",
    archivePrefix = "arXiv",
    primaryClass = "astro-ph.HE",
    doi = "10.3847/1538-4357/ad09af",
    journal = "Astrophys. J.",
    volume = "960",
    number = "1",
    pages = "82",
    year = "2024"
}

@article{Jacquemin-Ide:2023aax,
    author = "Jacquemin-Ide, Jonatan and Gottlieb, Ore and Lowell, Beverly and Tchekhovskoy, Alexander",
    title = "{Collapsar Gamma-Ray Bursts Grind Their Black Hole Spins to a Halt}",
    eprint = "2302.07281",
    archivePrefix = "arXiv",
    primaryClass = "astro-ph.HE",
    doi = "10.3847/1538-4357/ad02f0",
    journal = "Astrophys. J.",
    volume = "961",
    number = "2",
    pages = "212",
    year = "2024"
}

@article{Alexander:2008wv,
    author = "Alexander, R. D. and Armitage, P. J. and Cuadra, J.",
    title = "{Binary formation and mass function variations in fragmenting discs with short cooling times}",
    eprint = "0807.1731",
    archivePrefix = "arXiv",
    primaryClass = "astro-ph",
    doi = "10.1111/j.1365-2966.2008.13706.x",
    journal = "Mon. Not. R. Astron. Soc.",
    volume = "389",
    pages = "1655",
    year = "2008"
}

@article{Chandrasekhar:1931ih,
    author = "Chandrasekhar, Subrahmanyan",
    title = "{The maximum mass of ideal white dwarfs}",
    doi = "10.1086/143324",
    journal = "Astrophys. J.",
    volume = "74",
    pages = "81--82",
    year = "1931"
}

@article{Woosley:2020mze,
    author = "Woosley, S. and Sukhbold, Tuguldur and Janka, H. -T.",
    title = "{The Birth Function for Black Holes and Neutron Stars in Close Binaries}",
    eprint = "2001.10492",
    archivePrefix = "arXiv",
    primaryClass = "astro-ph.HE",
    doi = "10.3847/1538-4357/ab8cc1",
    journal = "Astrophys. J.",
    volume = "896",
    number = "1",
    pages = "56",
    year = "2020"
}

@article{Woosley:2006fn,
    author = "Woosley, S. E. and Bloom, J. S.",
    title = "{The Supernova Gamma-Ray Burst Connection}",
    eprint = "astro-ph/0609142",
    archivePrefix = "arXiv",
    reportNumber = "UCSC-ASTRO-2006-1",
    doi = "10.1146/annurev.astro.43.072103.150558",
    journal = "Annu. Rev. Astron. Astrophys.",
    volume = "44",
    pages = "507--556",
    year = "2006"
}

@article{Gottlieb:2023cgm,
    author = "Gottlieb, Ore and Jacquemin-Ide, Jonatan and Lowell, Beverly and Tchekhovskoy, Alexander and Ramirez-Ruiz, Enrico",
    title = "{Collapsar Black Holes Are Likely Born Slowly Spinning}",
    eprint = "2302.07271",
    archivePrefix = "arXiv",
    primaryClass = "astro-ph.HE",
    doi = "10.3847/2041-8213/ace779",
    journal = "Astrophys. J. Lett.",
    volume = "952",
    number = "2",
    pages = "L32",
    year = "2023"
}

@article{LVK:2022ydq,
    author = "Abbott, R. and others",
    collaboration = "LVK",
    title = "{Search for subsolar-mass black hole binaries in the second part of Advanced LIGO{\textquoteright}s and Advanced Virgo{\textquoteright}s third observing run}",
    eprint = "2212.01477",
    archivePrefix = "arXiv",
    primaryClass = "astro-ph.HE",
    doi = "10.1093/mnras/stad588",
    journal = "Mon. Not. R. Astron. Soc.",
    volume = "524",
    number = "4",
    pages = "5984--5992",
    year = "2023",
    note = "[Erratum: Mon.Not.Roy.Astron.Soc. 526, 6234 (2023)]"
}

@article{Kacanja:2024hme,
    author = "Kacanja, Keisi and Nitz, Alexander H.",
    title = "{A Search for Low-mass Neutron Stars in the Third Observing Run of Advanced LIGO and Virgo}",
    eprint = "2412.05369",
    archivePrefix = "arXiv",
    primaryClass = "astro-ph.HE",
    doi = "10.3847/1538-4357/adc9a5",
    journal = "Astrophys. J.",
    volume = "984",
    number = "1",
    pages = "61",
    year = "2025"
}

@article{Soni:2024cdb,
    author = "Soni, Kanchan and Nitz, Alexander H.",
    title = "{Hierarchical Searches for Subsolar-mass Binaries and the Third-generation Gravitational Wave Detector Era}",
    eprint = "2409.11317",
    archivePrefix = "arXiv",
    primaryClass = "gr-qc",
    doi = "10.3847/1538-4357/ad96b4",
    journal = "Astrophys. J.",
    volume = "978",
    number = "1",
    pages = "69",
    year = "2025"
}

@article{Vieira:2026eof,
    author = "Vieira, Nicholas and others",
    title = "{Search for a Counterpart to the Subsolar Mass Gravitational-wave Candidate S251112cm}",
    eprint = "2603.17009",
    archivePrefix = "arXiv",
    primaryClass = "astro-ph.HE",
    doi = "10.3847/1538-4357/ae877c",
    journal = "Astrophys. J.",
    volume = "1007",
    number = "2",
    pages = "116",
    year = "2026"
}

@article{Nitz:2020bdb,
    author = "Nitz, Alexander Harvey and Wang, Yi-Fan",
    title = "{Search for Gravitational Waves from High-Mass-Ratio Compact-Binary Mergers of Stellar Mass and Subsolar Mass Black Holes}",
    eprint = "2007.03583",
    archivePrefix = "arXiv",
    primaryClass = "astro-ph.HE",
    doi = "10.1103/PhysRevLett.126.021103",
    journal = "Phys. Rev. Lett.",
    volume = "126",
    number = "2",
    pages = "021103",
    year = "2021"
}

@article{Krnjaic:2026sgl,
    author = "Krnjaic, Gordan and Rocha, Duncan and Xiao, Huangyu",
    title = "{Primordial Neutron Stars}",
    eprint = "2604.08651",
    archivePrefix = "arXiv",
    primaryClass = "astro-ph.CO",
    reportNumber = "FERMILAB-PUB-26-0191-T",
    month = "4",
    year = "2026"
}

@book{Byrnes:2025tji,
    editor = "Byrnes, Christian and Franciolini, Gabriele and Harada, Tomohiro and Pani, Paolo and Sasaki, Misao",
    title = "{Primordial Black Holes}",
    doi = "10.1007/978-981-97-8887-3",
    isbn = "978-981--978886-6, 978-981--978889-7, 978-981--978887-3",
    publisher = "Springer",
    series = "Springer Series in Astrophysics and Cosmology",
    year = "2025"
}

@article{Gillanders:2025fwf,
    author = "Gillanders, J. H. and others",
    title = "{Pan-STARRS Follow-up of the Gravitational-wave Event S250818k and the Light Curve of SN2025ulz}",
    eprint = "2510.01142",
    archivePrefix = "arXiv",
    primaryClass = "astro-ph.HE",
    doi = "10.3847/2041-8213/ae2125",
    journal = "Astrophys. J. Lett.",
    volume = "995",
    number = "1",
    pages = "L27",
    year = "2025"
}

@article{Yang:2025dtj,
    author = "Yang, Yu-Han and others",
    title = "{AT2025ulz and S250818k: Zooming in with the Hubble Space Telescope}",
    eprint = "2510.18854",
    archivePrefix = "arXiv",
    primaryClass = "astro-ph.HE",
    doi = "10.3847/2041-8213/ae2dea",
    journal = "Astrophys. J. Lett.",
    volume = "996",
    number = "2",
    pages = "L24",
    year = "2026"
}

@article{Franz:2025uan,
    author = "Franz, Noah and others",
    title = "{Optimizing Kilonova Searches: A Case Study of the Type IIb SN 2025ulz in the Localization Volume of the Low-significance Gravitational Wave Event S250818k}",
    eprint = "2510.17104",
    archivePrefix = "arXiv",
    primaryClass = "astro-ph.HE",
    doi = "10.3847/2041-8213/ae17a8",
    journal = "Astrophys. J. Lett.",
    volume = "994",
    number = "2",
    pages = "L45",
    year = "2025"
}

@article{Piro:2006ja,
    author = "Piro, Anthony L. and Pfahl, Eric",
    title = "{Fragmentation of Collapsar Disks and the Production of Gravitational Waves}",
    eprint = "astro-ph/0610696",
    archivePrefix = "arXiv",
    doi = "10.1086/511672",
    journal = "Astrophys. J.",
    volume = "658",
    pages = "1173",
    year = "2007"
}

@article{Kasliwal:2025keb,
    author = "Kasliwal, Mansi M. and others",
    title = "{ZTF25abjmnps (AT2025ulz) and S250818k: A Candidate Superkilonova from a Subthreshold Subsolar Gravitational-wave Trigger}",
    eprint = "2510.23732",
    archivePrefix = "arXiv",
    primaryClass = "astro-ph.HE",
    doi = "10.3847/2041-8213/ae2000",
    journal = "Astrophys. J. Lett.",
    volume = "995",
    number = "2",
    pages = "L59",
    year = "2025"
}

@article{Lerner:2025dkd,
    author = "Lerner, Yonatan and Stone, Nicholas C. and Ofengeim, Dmitry D.",
    title = "{Fragmentation in collapsar discs: migration, growth, and emission}",
    eprint = "2505.21617",
    archivePrefix = "arXiv",
    primaryClass = "astro-ph.HE",
    doi = "10.1093/mnras/staf1835",
    journal = "Mon. Not. R. Astron. Soc.",
    volume = "545",
    number = "2",
    pages = "staf1835",
    year = "2025"
}

@article{Allen:2004gu,
    author = "Allen, Bruce",
    title = "{${\chi}^{2}$ time-frequency discriminator for gravitational wave detection}",
    eprint = "gr-qc/0405045",
    archivePrefix = "arXiv",
    doi = "10.1103/PhysRevD.71.062001",
    journal = "Phys. Rev. D",
    volume = "71",
    pages = "062001",
    year = "2005"
}

@article{Usman:2015kfa,
    author = "Usman, Samantha A. and others",
    title = "{The PyCBC search for gravitational waves from compact binary coalescence}",
    eprint = "1508.02357",
    archivePrefix = "arXiv",
    primaryClass = "gr-qc",
    reportNumber = "LIGO-P1500086",
    doi = "10.1088/0264-9381/33/21/215004",
    journal = "Class. Quant. Grav.",
    volume = "33",
    number = "21",
    pages = "215004",
    year = "2016"
}

@article{Cannon:2015gha,
    author = "Cannon, Kipp and Hanna, Chad and Peoples, Jacob",
    title = "{Likelihood-Ratio Ranking Statistic for Compact Binary Coalescence Candidates with Rate Estimation}",
    eprint = "1504.04632",
    archivePrefix = "arXiv",
    primaryClass = "astro-ph.IM",
    reportNumber = "LIGO-P1400175",
    month = "4",
    year = "2015"
}

@article{DalCanton:2014hxh,
    author = "Dal Canton, Tito and others",
    title = "{Implementing a search for aligned-spin neutron star-black hole systems with advanced ground based gravitational wave detectors}",
    eprint = "1405.6731",
    archivePrefix = "arXiv",
    primaryClass = "gr-qc",
    reportNumber = "LIGO-P1400053",
    doi = "10.1103/PhysRevD.90.082004",
    journal = "Phys. Rev. D",
    volume = "90",
    number = "8",
    pages = "082004",
    year = "2014"
}

@article{Nitz:2017svb,
    author = "Nitz, Alexander H. and Dent, Thomas and Dal Canton, Tito and Fairhurst, Stephen and Brown, Duncan A.",
    title = "{Detecting binary compact-object mergers with gravitational waves: Understanding and Improving the sensitivity of the PyCBC search}",
    eprint = "1705.01513",
    archivePrefix = "arXiv",
    primaryClass = "gr-qc",
    reportNumber = "LIGO-P1700088-V4",
    doi = "10.3847/1538-4357/aa8f50",
    journal = "Astrophys. J.",
    volume = "849",
    number = "2",
    pages = "118",
    year = "2017"
}

@article{Davies:2020tsx,
    author = "Davies, Gareth S. and Dent, Thomas and T{\'a}pai, M{\'a}rton and Harry, Ian and McIsaac, Connor and Nitz, Alexander H.",
    title = "{Extending the PyCBC search for gravitational waves from compact binary mergers to a global network}",
    eprint = "2002.08291",
    archivePrefix = "arXiv",
    primaryClass = "astro-ph.HE",
    doi = "10.1103/PhysRevD.102.022004",
    journal = "Phys. Rev. D",
    volume = "102",
    number = "2",
    pages = "022004",
    year = "2020"
}

@article{Tsukada:2023edh,
    author = "Tsukada, Leo and others",
    title = "{Improved ranking statistics of the GstLAL inspiral search for compact binary coalescences}",
    eprint = "2305.06286",
    archivePrefix = "arXiv",
    primaryClass = "astro-ph.IM",
    doi = "10.1103/PhysRevD.108.043004",
    journal = "Phys. Rev. D",
    volume = "108",
    number = "4",
    pages = "043004",
    year = "2023"
}

@article{ligo_scientific_collaboration_ligovirgokagra_2025,
    title = {{LIGO}/{Virgo}/{KAGRA} {S251112cm}: {Identification} of a {GW} compact binary merger candidate},
    volume = {42650},
    shorttitle = {{LIGO}/{Virgo}/{KAGRA} {S251112cm}},
    url = {https://ui.adsabs.harvard.edu/abs/2025GCN.42650....1L},
    urldate = {2026-02-12},
    journal = {GRB Coordinates Network},
    author = {{Ligo Scientific Collaboration} and {VIRGO Collaboration} and {Kagra Collaboration}},
    month = nov,
    year = {2025},
    pages = {1},
}

@article{Hall:2026gov,
    author = "Hall, Xander J. and others",
    title = "{Electromagnetic Follow-up of the Sub-Solar Mass Gravitational Wave Candidate S251112cm: Kilonova Constraints and a Coincident IIb Supernova}",
    eprint = "2605.10940",
    archivePrefix = "arXiv",
    primaryClass = "astro-ph.HE",
    month = "5",
    year = "2026"
}

@article{LIGOScientific:2016vlm,
    author = "Abbott, B. P. and others",
    collaboration = "LIGO Scientific, Virgo",
    title = "{Properties of the Binary Black Hole Merger GW150914}",
    eprint = "1602.03840",
    archivePrefix = "arXiv",
    primaryClass = "gr-qc",
    reportNumber = "LIGO-P1500218",
    doi = "10.1103/PhysRevLett.116.241102",
    journal = "Phys. Rev. Lett.",
    volume = "116",
    number = "24",
    pages = "241102",
    year = "2016"
}

@article{LIGOScientific:2025hdt,
    author = "Abac, A. G. and others",
    collaboration = "LIGO Scientific, KAGRA, VIRGO",
    title = "{GWTC-4.0: An Introduction to Version 4.0 of the Gravitational-Wave Transient Catalog}",
    eprint = "2508.18080",
    archivePrefix = "arXiv",
    primaryClass = "gr-qc",
    reportNumber = "LIGO-P2400293",
    doi = "10.3847/2041-8213/ae0c06",
    journal = "Astrophys. J. Lett.",
    volume = "995",
    number = "1",
    pages = "L18",
    year = "2025"
}

@article{Singh:2020wiq,
    author = "Singh, Divya and Ryan, Michael and Magee, Ryan and Akhter, Towsifa and Shandera, Sarah and Jeong, Donghui and Hanna, Chad",
    title = "{Gravitational-wave limit on the Chandrasekhar mass of dark matter}",
    eprint = "2009.05209",
    archivePrefix = "arXiv",
    primaryClass = "astro-ph.CO",
    reportNumber = "LIGO-P2000291",
    doi = "10.1103/PhysRevD.104.044015",
    journal = "Phys. Rev. D",
    volume = "104",
    number = "4",
    pages = "044015",
    year = "2021"
}

@article{Siegel:2018zxq,
    author = "Siegel, Daniel M. and Barnes, Jennifer and Metzger, Brian D.",
    title = "{Collapsars as a major source of r-process elements}",
    eprint = "1810.00098",
    archivePrefix = "arXiv",
    primaryClass = "astro-ph.HE",
    doi = "10.1038/s41586-019-1136-0",
    journal = "Nature",
    volume = "569",
    pages = "241",
    year = "2019"
}

@article{Haensel:1986qb,
    author = "Haensel, P. and Zdunik, J. L. and Schaeffer, R.",
    title = "{Strange quark stars}",
    journal = "Astron. Astrophys.",
    volume = "160",
    pages = "121--128",
    year = "1986"
}

@article{Jaranowski:1994xd,
    author = "Jaranowski, P. and Krolak, A.",
    title = "{Optimal solution to the inverse problem for the gravitational wave signal of a coalescing compact binary}",
    doi = "10.1103/PhysRevD.49.1723",
    journal = "Phys. Rev. D",
    volume = "49",
    pages = "1723--1739",
    year = "1994"
}

@article{Blanchet:2013haa,
    author = "Blanchet, Luc",
    title = "{Post-Newtonian Theory for Gravitational Waves}",
    eprint = "1310.1528",
    archivePrefix = "arXiv",
    primaryClass = "gr-qc",
    doi = "10.12942/lrr-2014-2",
    journal = "Living Rev. Relativ.",
    volume = "17",
    pages = "2",
    year = "2014"
}

@article{Prakash:1996xs,
    author = "Prakash, Madappa and Bombaci, Ignazio and Prakash, Manju and Ellis, Paul J. and Lattimer, James M. and Knorren, Roland",
    title = "{Composition and structure of protoneutron stars}",
    eprint = "nucl-th/9603042",
    archivePrefix = "arXiv",
    reportNumber = "SUNY-NTG-96-11, NUC-MINN-93-23-T",
    doi = "10.1016/S0370-1573(96)00023-3",
    journal = "Phys. Rep.",
    volume = "280",
    pages = "1--77",
    year = "1997"
}

@article{Prakash:1995uw,
    author = "Prakash, M. and Cooke, J. R. and Lattimer, J. M.",
    title = "{Quark - hadron phase transition in protoneutron stars}",
    doi = "10.1103/PhysRevD.52.661",
    journal = "Phys. Rev. D",
    volume = "52",
    pages = "661--665",
    year = "1995"
}

@article{Gondek:1997fd,
    author = "Gondek, D. and Haensel, P. and Zdunik, J. L.",
    title = "{Radial pulsations and stability of protoneutron stars}",
    eprint = "astro-ph/9705157",
    archivePrefix = "arXiv",
    journal = "Astron. Astrophys.",
    volume = "325",
    pages = "217--227",
    year = "1997"
}

@article{2017MNRAS.471.4917J,
       author = {{Jim{\'e}nez}, Mar{\'\i}a Alejandra and {Masset}, Fr{\'e}d{\'e}ric S.},
        title = "{Improved torque formula for low- and intermediate-mass planetary migration}",
      journal = {Monthly Notices of the Royal Astronomical Society},
         year = 2017,
        month = nov,
       volume = {471},
       number = {4},
        pages = {4917-4929},
          doi = {10.1093/mnras/stx1946},
archivePrefix = {arXiv},
       eprint = {1707.08988},
 primaryClass = {astro-ph.EP},
       adsurl = {https://ui.adsabs.harvard.edu/abs/2017MNRAS.471.4917J}
}

@article{1977MNRAS.179..433B,
       author = {{Blandford}, R.~D. and {Znajek}, R.~L.},
        title = "{Electromagnetic extraction of energy from Kerr black holes.}",
      journal = {Monthly Notices of the Royal Astronomical Society},
         year = 1977,
        month = may,
       volume = {179},
        pages = {433-456},
          doi = {10.1093/mnras/179.3.433},
       adsurl = {https://ui.adsabs.harvard.edu/abs/1977MNRAS.179..433B}
}

@article{Essick:2024olf,
    author = "Essick, Reed",
    title = "{Exotic Stable Branches with Efficient TOV Sequences}",
    eprint = "2405.05395",
    archivePrefix = "arXiv",
    primaryClass = "astro-ph.HE",
    doi = "10.3847/2041-8213/ad7399",
    journal = "Astrophys. J. Lett.",
    volume = "973",
    number = "2",
    pages = "L50",
    year = "2024"
}

@article{Janka:2025tvf,
    author = "Janka, Hans-Thomas",
    title = "{Long-Term Multidimensional Models of Core-Collapse Supernovae: Progress and Challenges}",
    eprint = "2502.14836",
    archivePrefix = "arXiv",
    primaryClass = "astro-ph.HE",
    doi = "10.1146/annurev-nucl-121423-100945",
    journal = "Ann. Rev. Nucl. Part. Sci.",
    volume = "75",
    number = "1",
    pages = "425--461",
    year = "2025"
}

@article{Burrows:2020qrp,
    author = "Burrows, Adam and Vartanyan, David",
    title = "{Core-Collapse Supernova Explosion Theory}",
    eprint = "2009.14157",
    archivePrefix = "arXiv",
    primaryClass = "astro-ph.SR",
    doi = "10.1038/s41586-020-03059-w",
    journal = "Nature",
    volume = "589",
    number = "7840",
    pages = "29--39",
    year = "2021"
}

@article{Lattimer:2021emm,
    author = "Lattimer, J. M.",
    title = "{Neutron Stars and the Nuclear Matter Equation of State}",
    doi = "10.1146/annurev-nucl-102419-124827",
    journal = "Ann. Rev. Nucl. Part. Sci.",
    volume = "71",
    pages = "433--464",
    year = "2021"
}

@article{Schneider:2020vvh,
    author = {Schneider, F. R. N. and Podsiadlowski, Ph. and M{\"u}ller, B.},
    title = "{Pre-supernova evolution, compact object masses and explosion properties of stripped binary stars}",
    eprint = "2008.08599",
    archivePrefix = "arXiv",
    primaryClass = "astro-ph.SR",
    doi = "10.1051/0004-6361/202039219",
    journal = "Astron. Astrophys.",
    volume = "645",
    pages = "A5",
    year = "2021"
}

@article{Russo:2025azm,
    author = "Russo, Benedetta and Urbano, Alfredo",
    title = "{Tidal gap: Causality bound on exotic compact objects}",
    eprint = "2512.19519",
    archivePrefix = "arXiv",
    primaryClass = "gr-qc",
    doi = "10.1103/wsc5-m2w6",
    journal = "Phys. Rev. D",
    volume = "114",
    number = "10",
    pages = "104007",
    year = "2026"
}

@article{LIGOScientific:2026wxz,
    author = "Abac, A. G. and others",
    collaboration = "LIGO Scientific, VIRGO, KAGRA",
    title = "{Searches for Binary Mergers with Sub-solar Mass Components in Data from the First Part of LIGO--Virgo--KAGRA's Fourth Observing Run}",
    eprint = "2605.05444",
    archivePrefix = "arXiv",
    primaryClass = "astro-ph.HE",
    month = "5",
    year = "2026"
}

@article{2025GCN.41437....1L,
       author = {{Ligo Scientific Collaboration} and {VIRGO Collaboration} and {Kagra Collaboration}},
        title = "{LIGO/Virgo/KAGRA S250818k: Properties of the low-significance GW compact binary merger candidate potentially associated with AT 2025ulz}",
      journal = {GRB Coordinates Network},
         year = 2025,
        month = aug,
       volume = {41437},
        pages = {1},
       adsurl = {https://ui.adsabs.harvard.edu/abs/2025GCN.41437....1L}
}

@article{Paek:2026kat,
    author = "Paek, Gregory S. H. and Olivares E., Felipe and Hoogendam, Willem B.",
    title = "{A 2 month Galaxy-targeted Near-infrared Follow-up of the Sub-solar-mass Gravitational-wave Candidate S251112cm: Probing Electromagnetic Counterpart Scenarios}",
    eprint = "2608.22775",
    archivePrefix = "arXiv",
    primaryClass = "astro-ph.HE",
    doi = "10.3847/1538-4357/ae9e26",
    journal = "Astrophys. J.",
    volume = "1010",
    number = "2",
    pages = "166",
    year = "2026"
}

@article{Crescimbeni:2024qrq,
    author = "Crescimbeni, Francesco and Franciolini, Gabriele and Pani, Paolo and Vaglio, Massimo",
    title = "{Cosmology and nuclear physics implications of a subsolar gravitational-wave event}",
    eprint = "2408.14287",
    archivePrefix = "arXiv",
    primaryClass = "astro-ph.HE",
    reportNumber = "CERN-TH-2024-121",
    doi = "10.1103/PhysRevD.111.083538",
    journal = "Phys. Rev. D",
    volume = "111",
    number = "8",
    pages = "083538",
    year = "2025"
}

@article{MacFadyen:1998vz,
    author = "MacFadyen, A. and Woosley, S. E.",
    title = "{Collapsars: Gamma-ray bursts and explosions in 'failed supernovae'}",
    eprint = "astro-ph/9810274",
    archivePrefix = "arXiv",
    doi = "10.1086/307790",
    journal = "Astrophys. J.",
    volume = "524",
    pages = "262",
    year = "1999"
}

@article{Sukhbold:2015wba,
    author = "Sukhbold, Tuguldur and Ertl, T. and Woosley, S. E. and Brown, Justin M. and Janka, H. -T.",
    title = "{Core-Collapse Supernovae from 9 to 120 Solar Masses Based on Neutrino-powered Explosions}",
    eprint = "1510.04643",
    archivePrefix = "arXiv",
    primaryClass = "astro-ph.HE",
    doi = "10.3847/0004-637X/821/1/38",
    journal = "Astrophys. J.",
    volume = "821",
    number = "1",
    pages = "38",
    year = "2016"
}

@article{Ertl:2019zks,
    author = "Ertl, Thomas and Woosley, Stan E. and Sukhbold, Tuguldur and Janka, H. -Thomas",
    title = "{The Explosion of Helium Stars Evolved With Mass Loss}",
    eprint = "1910.01641",
    archivePrefix = "arXiv",
    primaryClass = "astro-ph.HE",
    doi = "10.3847/1538-4357/ab6458",
    month = "10",
    year = "2019"
}

@article{Burrows:2019rtd,
    author = "Burrows, Adam and Radice, David and Vartanyan, David",
    title = "{Three-dimensional supernova explosion simulations of 9-, 10-, 11-, 12-, and 13-M{\ensuremath{\odot}} stars}",
    eprint = "1902.00547",
    archivePrefix = "arXiv",
    primaryClass = "astro-ph.SR",
    doi = "10.1093/mnras/stz543",
    journal = "Mon. Not. R. Astron. Soc.",
    volume = "485",
    number = "3",
    pages = "3153--3168",
    year = "2019"
}

@article{CanevaSantoro:2023aol,
    author = "Caneva Santoro, Giada and Roy, Soumen and Vicente, Rodrigo and Haney, Maria and Piccinni, Ornella Juliana and Del Pozzo, Walter and Martinez, Mario",
    title = "{First Constraints on Compact Binary Environments from LIGO-Virgo Data}",
    eprint = "2309.05061",
    archivePrefix = "arXiv",
    primaryClass = "gr-qc",
    reportNumber = "LIGO DCC P2300301",
    doi = "10.1103/PhysRevLett.132.251401",
    journal = "Phys. Rev. Lett.",
    volume = "132",
    number = "25",
    pages = "251401",
    year = "2024"
}

@article{Metzger:2024ujc,
    author = "Metzger, Brian D. and Hui, Lam and Cantiello, Matteo",
    title = "{Fragmentation in Gravitationally Unstable Collapsar Disks and Subsolar Neutron Star Mergers}",
    eprint = "2407.07955",
    archivePrefix = "arXiv",
    primaryClass = "astro-ph.HE",
    doi = "10.3847/2041-8213/ad6990",
    journal = "Astrophys. J. Lett.",
    volume = "971",
    number = "2",
    pages = "L34",
    year = "2024"
}

@article{Tejera:2026dpr,
    author = "Tejera, Abby and Divyajyoti and Corsi, Alessandra and Zenati, Yossef and Coyne, Robert and Cote, Logan",
    title = "{Long-duration GW Searches for Sub-solar NSs and Superkilonovae using CoCoA}",
    eprint = "2608.11387",
    archivePrefix = "arXiv",
    primaryClass = "astro-ph.HE",
    month = "8",
    year = "2026"
}

@article{Allen:2005fk,
    author = "Allen, Bruce and Anderson, Warren G. and Brady, Patrick R. and Brown, Duncan A. and Creighton, Jolien D. E.",
    title = "{FINDCHIRP: An Algorithm for detection of gravitational waves from inspiraling compact binaries}",
    eprint = "gr-qc/0509116",
    archivePrefix = "arXiv",
    doi = "10.1103/PhysRevD.85.122006",
    journal = "Phys. Rev. D",
    volume = "85",
    pages = "122006",
    year = "2012"
}

@article{Zeldovich:1967lct,
	title        = {{The Hypothesis of Cores Retarded during Expansion and the Hot Cosmological Model}},
	author       = {Zel'dovich, Ya. B. and Novikov, I. D.},
	year         = 1967,
	journal      = {Sov. Astron.},
	volume       = 10,
	pages        = 602
}

@article{Hawking:1971ei,
	title        = {{Gravitationally collapsed objects of very low mass}},
	author       = {Hawking, Stephen},
	year         = 1971,
	journal      = {Mon. Not. R. Astron. Soc.},
	volume       = 152,
	pages        = 75,
	doi          = {10.1093/mnras/152.1.75}
}

@article{Carr:1974nx,
	title        = {{Black holes in the early Universe}},
	author       = {Carr, Bernard J. and Hawking, S. W.},
	year         = 1974,
	journal      = {Mon. Not. R. Astron. Soc.},
	volume       = 168,
	pages        = {399--415},
	doi          = {10.1093/mnras/168.2.399}
}

@article{Carr:1975qj,
	title        = {{The Primordial black hole mass spectrum}},
	author       = {Carr, Bernard J.},
	year         = 1975,
	journal      = {Astrophys. J.},
	volume       = 201,
	pages        = {1--19},
	doi          = {10.1086/153853}
}

@article{Sasaki:2016jop,
	title        = {{Primordial Black Hole Scenario for the Gravitational-Wave Event GW150914}},
	author       = {Sasaki, Misao and Suyama, Teruaki and Tanaka, Takahiro and Yokoyama, Shuichiro},
	year         = 2016,
	journal      = {Phys. Rev. Lett.},
	volume       = 117,
	number       = 6,
	pages        = {061101},
	doi          = {10.1103/PhysRevLett.117.061101},
	note         = {[Erratum: Phys.Rev.Lett. 121, 059901 (2018)]},
	eprint       = {1603.08338},
	archiveprefix = {arXiv},
	primaryclass = {astro-ph.CO},
	reportnumber = {RESCEU-17-16, RUP-16-7, YITP-16-43}
}

@article{DeLuca:2024uju,
    author = "De Luca, Valerio and Franciolini, Gabriele and Riotto, Antonio",
    title = "{Flea on the elephant: Tidal Love numbers in subsolar primordial black hole searches}",
    eprint = "2408.14207",
    archivePrefix = "arXiv",
    primaryClass = "gr-qc",
    reportNumber = "CERN-TH-2024-140",
    doi = "10.1103/PhysRevD.110.104041",
    journal = "Phys. Rev. D",
    volume = "110",
    number = "10",
    pages = "104041",
    year = "2024"
}

@article{Raidal:2018bbj,
	title        = {{Formation and Evolution of Primordial Black Hole Binaries in the Early Universe}},
	author       = {Raidal, Martti and Spethmann, Christian and Vaskonen, Ville and Veerm{\"a}e, Hardi},
	year         = 2019,
	journal      = {J. Cosmology Astropart. Phys.},
	volume       = {02},
	pages        = {018},
	doi          = {10.1088/1475-7516/2019/02/018},
	eprint       = {1812.01930},
	archiveprefix = {arXiv},
	primaryclass = {astro-ph.CO},
	reportnumber = {KCL-PH-TH/2018-70, CERN-TH-2018-266}
}

@article{Bird:2016dcv,
	title        = {{Did LIGO detect dark matter?}},
	author       = {Bird, Simeon and Cholis, Ilias and Mu{\~n}oz, Julian B. and Ali-Ha{\"\i}moud, Yacine and Kamionkowski, Marc and Kovetz, Ely D. and Raccanelli, Alvise and Riess, Adam G.},
	year         = 2016,
	journal      = {Phys. Rev. Lett.},
	volume       = 116,
	number       = 20,
	pages        = 201301,
	doi          = {10.1103/PhysRevLett.116.201301},
	eprint       = {1603.00464},
	archiveprefix = {arXiv},
	primaryclass = {astro-ph.CO}
}

@article{Clesse:2016vqa,
	title        = {{The clustering of massive Primordial Black Holes as Dark Matter: measuring their mass distribution with Advanced LIGO}},
	author       = {Clesse, Sebastien and Garc{\'\i}a-Bellido, Juan},
	year         = 2017,
	journal      = {Phys. Dark Univ.},
	volume       = 15,
	pages        = {142--147},
	doi          = {10.1016/j.dark.2016.10.002},
	eprint       = {1603.05234},
	archiveprefix = {arXiv},
	primaryclass = {astro-ph.CO},
	reportnumber = {TTK-16-10, IFT-UAM-CSIC-16-027}
}

@article{LISACosmologyWorkingGroup:2023njw,
	title        = {{Primordial black holes and their gravitational-wave signatures}},
	author       = {Bagui, Eleni and others},
	year         = 2025,
	journal      = {Living Rev. Relativ.},
	volume       = 28,
	number       = 1,
	pages        = 1,
	doi          = {10.1007/s41114-024-00053-w},
	collaboration = {LISA Cosmology Working Group},
	eprint       = {2310.19857},
	archiveprefix = {arXiv},
	primaryclass = {astro-ph.CO}
}

@article{Franciolini:2021xbq,
	title        = {{How to assess the primordial origin of single gravitational-wave events with mass, spin, eccentricity, and deformability measurements}},
	author       = {Franciolini, Gabriele and Cotesta, Roberto and Loutrel, Nicholas and Berti, Emanuele and Pani, Paolo and Riotto, Antonio},
	year         = 2022,
	journal      = {Phys. Rev. D},
	volume       = 105,
	number       = 6,
	pages        = {063510},
	doi          = {10.1103/PhysRevD.105.063510},
	eprint       = {2112.10660},
	archiveprefix = {arXiv},
	primaryclass = {astro-ph.CO},
	reportnumber = {ET-0464A-21}
}

@inbook{Raidal:2024bmm,
	title        = {{Formation of~Primordial Black Hole Binaries and~Their Merger Rates}},
	author       = {Raidal, Martti and Vaskonen, Ville and Veerm{\"a}e, Hardi},
	year         = 2025,
	booktitle    = {{Primordial Black Holes}},
	publisher    = {Springer},
	doi          = {10.1007/978-981-97-8887-3_16},
	editor       = {Byrnes, Christian and Franciolini, Gabriele and Harada, Tomohiro and Pani, Paolo and Sasaki, Misao},
	eprint       = {2404.08416},
	archiveprefix = {arXiv},
	primaryclass = {astro-ph.CO}
}

@article{Peters:1964zz,
	title        = {{Gravitational Radiation and the Motion of Two Point Masses}},
	author       = {Peters, P. C.},
	year         = 1964,
	journal      = {Phys. Rev.},
	volume       = 136,
	pages        = {B1224--B1232},
	doi          = {10.1103/PhysRev.136.B1224}
}

@article{Peters:1963ux,
	title        = {{Gravitational radiation from point masses in a Keplerian orbit}},
	author       = {Peters, P. C. and Mathews, J.},
	year         = 1963,
	journal      = {Phys. Rev.},
	volume       = 131,
	pages        = {435--439},
	doi          = {10.1103/PhysRev.131.435}
}

@article{Carr:2026hot,
    author = {Carr, Bernard and Iovino, Antonio J. and Perna, Gabriele and Vaskonen, Ville and Veerm{\"a}e, Hardi},
    title = "{Primordial black holes: constraints, potential evidence and prospects}",
    eprint = "2601.06024",
    archivePrefix = "arXiv",
    primaryClass = "astro-ph.CO",
    doi = "10.1007/s40766-026-00080-z",
    journal = "Riv. Nuovo Cim.",
    volume = "49",
    number = "5",
    pages = "225--274",
    year = "2026"
}

@article{Wu:2023qeh,
    author = "Wu, Yinhao and Chen, Yi-Xian and Lin, Douglas N. C.",
    title = "{Chaotic Type I migration in turbulent discs}",
    eprint = "2311.15747",
    archivePrefix = "arXiv",
    primaryClass = "astro-ph.EP",
    doi = "10.1093/mnrasl/slad183",
    journal = "Mon. Not. R. Astron. Soc.",
    volume = "528",
    number = "1",
    pages = "L127--L132",
    year = "2023"
}

@article{Chen:2025uwd,
    author = "Chen, Yi-Xian and Metzger, Brian D.",
    title = "{Gravitational Instability and Fragmentation in Collapsar Disks Supports the Formation of Subsolar Neutron Stars}",
    eprint = "2508.17183",
    archivePrefix = "arXiv",
    primaryClass = "astro-ph.HE",
    doi = "10.3847/2041-8213/ae045d",
    journal = "Astrophys. J. Lett.",
    volume = "991",
    number = "1",
    pages = "L22",
    year = "2025"
}

@article{Nitz:2021mzz,
    author = "Nitz, Alexander H. and Wang, Yi-Fan",
    title = "{Search for gravitational waves from the coalescence of sub-solar mass and eccentric compact binaries}",
    eprint = "2102.00868",
    archivePrefix = "arXiv",
    primaryClass = "astro-ph.HE",
    doi = "10.3847/1538-4357/ac01d9",
    journal = "Astrophys. J.",
    volume = "915",
    pages = "54",
    month = "2",
    year = "2021"
}

@article{LIGOScientific:2021usb,
	title        = {{GWTC-2.1: Deep extended catalog of compact binary coalescences observed by LIGO and Virgo during the first half of the third observing run}},
	author       = {Abbott, R. and others},
	year         = 2024,
	journal      = {Phys. Rev. D},
	volume       = 109,
	number       = 2,
	pages        = {022001},
	doi          = {10.1103/PhysRevD.109.022001},
	collaboration = {LIGO Scientific, VIRGO},
	eprint       = {2108.01045},
	archiveprefix = {arXiv},
	primaryclass = {gr-qc},
	reportnumber = {LIGO-P2100063}
}

@article{Berti:2005ys,
	title        = {{On gravitational-wave spectroscopy of massive black holes with the space interferometer LISA}},
	author       = {Berti, Emanuele and Cardoso, Vitor and Will, Clifford M.},
	year         = 2006,
	journal      = {Phys. Rev. D},
	volume       = 73,
	pages        = {064030},
	doi          = {10.1103/PhysRevD.73.064030},
	eprint       = {gr-qc/0512160},
	archiveprefix = {arXiv}
}

@article{Barsanti:2021ydd,
	title        = {{Detecting Subsolar-Mass Primordial Black Holes in Extreme Mass-Ratio Inspirals with LISA and Einstein Telescope}},
	author       = {Barsanti, Susanna and De Luca, Valerio and Maselli, Andrea and Pani, Paolo},
	year         = 2022,
	journal      = {Phys. Rev. Lett.},
	volume       = 128,
	number       = 11,
	pages        = 111104,
	doi          = {10.1103/PhysRevLett.128.111104},
	eprint       = {2109.02170},
	archiveprefix = {arXiv},
	primaryclass = {gr-qc}
}

@article{Franciolini:2021tla,
	title        = {{Searching for a subpopulation of primordial black holes in LIGO-Virgo gravitational-wave data}},
	author       = {Franciolini, Gabriele and Baibhav, Vishal and De Luca, Valerio and Ng, Ken K. Y. and Wong, Kaze W. K. and Berti, Emanuele and Pani, Paolo and Riotto, Antonio and Vitale, Salvatore},
	year         = 2022,
	journal      = {Phys. Rev. D},
	volume       = 105,
	number       = 8,
	pages        = {083526},
	doi          = {10.1103/PhysRevD.105.083526},
	eprint       = {2105.03349},
	archiveprefix = {arXiv},
	primaryclass = {gr-qc}
}

@article{Carr:2020gox,
	title        = {{Constraints on primordial black holes}},
	author       = {Carr, Bernard and Kohri, Kazunori and Sendouda, Yuuiti and Yokoyama, Jun'ichi},
	year         = 2021,
	journal      = {Rep. Prog. Phys.},
	volume       = 84,
	number       = 11,
	pages        = 116902,
	doi          = {10.1088/1361-6633/ac1e31},
	eprint       = {2002.12778},
	archiveprefix = {arXiv},
	primaryclass = {astro-ph.CO},
	reportnumber = {RESCEU-03/20; KEK-Cosmo-249; KEK-TH-2199; IPMU20-0024}
}

@article{Green:2020jor,
	title        = {{Primordial Black Holes as a dark matter candidate}},
	author       = {Green, Anne M. and Kavanagh, Bradley J.},
	year         = 2021,
	journal      = {J. Phys. G},
	volume       = 48,
	number       = 4,
	pages        = {043001},
	doi          = {10.1088/1361-6471/abc534},
	eprint       = {2007.10722},
	archiveprefix = {arXiv},
	primaryclass = {astro-ph.CO}
}

@article{Sasaki:2018dmp,
	title        = {{Primordial black holes\textemdash{}perspectives in gravitational wave astronomy}},
	author       = {Sasaki, Misao and Suyama, Teruaki and Tanaka, Takahiro and Yokoyama, Shuichiro},
	year         = 2018,
	journal      = {Class. Quantum Grav.},
	volume       = 35,
	number       = 6,
	pages        = {063001},
	doi          = {10.1088/1361-6382/aaa7b4},
	eprint       = {1801.05235},
	archiveprefix = {arXiv},
	primaryclass = {astro-ph.CO}
}

@article{LIGOScientific:2018mvr,
	title        = {{GWTC-1: A Gravitational-Wave Transient Catalog of Compact Binary Mergers Observed by LIGO and Virgo during the First and Second Observing Runs}},
	author       = {Abbott, B. P. and others},
	year         = 2019,
	journal      = {Phys. Rev. X},
	volume       = 9,
	number       = 3,
	pages        = {031040},
	doi          = {10.1103/PhysRevX.9.031040},
	collaboration = {LIGO Scientific, Virgo},
	eprint       = {1811.12907},
	archiveprefix = {arXiv},
	primaryclass = {astro-ph.HE},
	reportnumber = {LIGO-P1800307}
}

@article{LIGOScientific:2020ibl,
	title        = {{GWTC-2: Compact Binary Coalescences Observed by LIGO and Virgo During the First Half of the Third Observing Run}},
	author       = {Abbott, R. and others},
	year         = 2021,
	journal      = {Phys. Rev. X},
	volume       = 11,
	pages        = {021053},
	doi          = {10.1103/PhysRevX.11.021053},
	collaboration = {LIGO Scientific, Virgo},
	eprint       = {2010.14527},
	archiveprefix = {arXiv},
	primaryclass = {gr-qc},
	reportnumber = {P2000061}
}

@article{Crescimbeni:2024cwh,
	title        = {{Can we identify primordial black holes? Tidal tests for subsolar-mass gravitational-wave observations}},
	author       = {Crescimbeni, Francesco and Franciolini, Gabriele and Pani, Paolo and Riotto, Antonio},
	year         = 2024,
	journal      = {Phys. Rev. D},
	volume       = 109,
	number       = 12,
	pages        = 124063,
	doi          = {10.1103/PhysRevD.109.124063},
	eprint       = {2402.18656},
	archiveprefix = {arXiv},
	primaryclass = {astro-ph.HE},
	reportnumber = {CERN-TH-2024-026}
}

@article{Andres-Carcasona:2024wqk,
	title        = {{Constraints on primordial black holes from LIGO-Virgo-KAGRA O3 events}},
	author       = {Andr\'es-Carcasona, M. and Iovino, A. J. and Vaskonen, V. and Veerm\"ae, H. and Mart\'\i{}nez, M. and Pujol\`as, O. and Mir, Ll. M.},
	year         = 2024,
	journal      = {Phys. Rev. D},
	volume       = 110,
	number       = 2,
	pages        = {023040},
	doi          = {10.1103/PhysRevD.110.023040},
	eprint       = {2405.05732},
	archiveprefix = {arXiv},
	primaryclass = {astro-ph.CO}
}

@article{Hinderer:2007mb,
	title        = {{Tidal Love numbers of neutron stars}},
	author       = {Hinderer, Tanja},
	year         = 2008,
	journal      = {Astrophys. J.},
	volume       = 677,
	pages        = {1216--1220},
	doi          = {10.1086/533487},
	note         = {[Erratum: Astrophys.J. 697, 964 (2009)]},
	eprint       = {0711.2420},
	archiveprefix = {arXiv},
	primaryclass = {astro-ph}
}

@article{Flanagan:2007ix,
	title        = {{Constraining neutron star tidal Love numbers with gravitational wave detectors}},
	author       = {Flanagan, Eanna E. and Hinderer, Tanja},
	year         = 2008,
	journal      = {Phys. Rev. D},
	volume       = 77,
	pages        = {021502},
	doi          = {10.1103/PhysRevD.77.021502},
	eprint       = {0709.1915},
	archiveprefix = {arXiv},
	primaryclass = {astro-ph}
}

@article{Binnington:2009bb,
	title        = {{Relativistic theory of tidal Love numbers}},
	author       = {Binnington, Taylor and Poisson, Eric},
	year         = 2009,
	journal      = {Phys. Rev. D},
	volume       = 80,
	pages        = {084018},
	doi          = {10.1103/PhysRevD.80.084018},
	eprint       = {0906.1366},
	archiveprefix = {arXiv},
	primaryclass = {gr-qc}
}

@article{Cardoso:2019upw,
	title        = {{Environmental effects in gravitational-wave physics: Tidal deformability of black holes immersed in matter}},
	author       = {Cardoso, Vitor and Duque, Francisco},
	year         = 2020,
	journal      = {Phys. Rev. D},
	volume       = 101,
	number       = 6,
	pages        = {064028},
	doi          = {10.1103/PhysRevD.101.064028},
	eprint       = {1912.07616},
	archiveprefix = {arXiv},
	primaryclass = {gr-qc}
}

@article{DeLuca:2022xlz,
	title        = {{Modeling frequency-dependent tidal deformability for environmental black hole mergers}},
	author       = {De Luca, Valerio and Maselli, Andrea and Pani, Paolo},
	year         = 2023,
	journal      = {Phys. Rev. D},
	volume       = 107,
	number       = 4,
	pages        = {044058},
	doi          = {10.1103/PhysRevD.107.044058},
	eprint       = {2212.03343},
	archiveprefix = {arXiv},
	primaryclass = {gr-qc}
}

@article{Damour:2009va,
	title        = {{On the gravitational polarizability of black holes}},
	author       = {Damour, Thibault and Lecian, Orchidea Maria},
	year         = 2009,
	journal      = {Phys. Rev. D},
	volume       = 80,
	pages        = {044017},
	doi          = {10.1103/PhysRevD.80.044017},
	eprint       = {0906.3003},
	archiveprefix = {arXiv},
	primaryclass = {gr-qc},
	reportnumber = {IHES-P-09-28}
}

@article{Cardoso:2017cfl,
	title        = {{Testing strong-field gravity with tidal Love numbers}},
	author       = {Cardoso, Vitor and Franzin, Edgardo and Maselli, Andrea and Pani, Paolo and Raposo, Guilherme},
	year         = 2017,
	journal      = {Phys. Rev. D},
	volume       = 95,
	number       = 8,
	pages        = {084014},
	doi          = {10.1103/PhysRevD.95.084014},
	note         = {[Addendum: Phys.Rev.D 95, 089901 (2017)]},
	eprint       = {1701.01116},
	archiveprefix = {arXiv},
	primaryclass = {gr-qc}
}

@article{Hui:2020xxx,
	title        = {{Static response and Love numbers of Schwarzschild black holes}},
	author       = {Hui, Lam and Joyce, Austin and Penco, Riccardo and Santoni, Luca and Solomon, Adam R.},
	year         = 2021,
	journal      = {J. Cosmology Astropart. Phys.},
	volume       = {04},
	pages        = {052},
	doi          = {10.1088/1475-7516/2021/04/052},
	eprint       = {2010.00593},
	archiveprefix = {arXiv},
	primaryclass = {hep-th}
}

@article{Flanagan:1997sx,
	title        = {{Measuring gravitational waves from binary black hole coalescences: 1. Signal-to-noise for inspiral, merger, and ringdown}},
	author       = {Flanagan, Eanna E. and Hughes, Scott A.},
	year         = 1998,
	journal      = {Phys. Rev. D},
	volume       = 57,
	pages        = {4535--4565},
	doi          = {10.1103/PhysRevD.57.4535},
	eprint       = {gr-qc/9701039},
	archiveprefix = {arXiv},
	reportnumber = {GRP-456}
}

@article{Bandopadhyay:2022tbi,
	title        = {{Detectability of subsolar mass neutron stars through a template bank search}},
	author       = {Bandopadhyay, Ananya and Reed, Brendan and Padamata, Surendra and Leon, Erick and Horowitz, C. J. and Brown, Duncan A. and Radice, David and Fattoyev, F. J. and Piekarewicz, J.},
	year         = 2023,
	journal      = {Phys. Rev. D},
	volume       = 107,
	number       = 10,
	pages        = 103012,
	doi          = {10.1103/PhysRevD.107.103012},
	eprint       = {2212.03855},
	archiveprefix = {arXiv},
	primaryclass = {astro-ph.HE}
}

@article{Owen:1998dk,
	title        = {{Matched filtering of gravitational waves from inspiraling compact binaries: Computational cost and template placement}},
	author       = {Owen, Benjamin J. and Sathyaprakash, B. S.},
	year         = 1999,
	journal      = {Phys. Rev. D},
	volume       = 60,
	pages        = {022002},
	doi          = {10.1103/PhysRevD.60.022002},
	eprint       = {gr-qc/9808076},
	archiveprefix = {arXiv},
	reportnumber = {GRP-505}
}

\clearpage
\onecolumngrid

\begin{center}
  \textbf{\large Supplemental Material}
\end{center}

\setcounter{equation}{0}
\setcounter{figure}{0}
\setcounter{table}{0}
\setcounter{section}{0}
\renewcommand{\theequation}{S\arabic{equation}}
\renewcommand{\thefigure}{S\arabic{figure}}
\renewcommand{\thetable}{S\arabic{table}}

\twocolumngrid

\noindent
This Supplemental Material provides additional details supporting the results
of the main text.

\begin{figure*}[t]
\includegraphics[width=\linewidth]{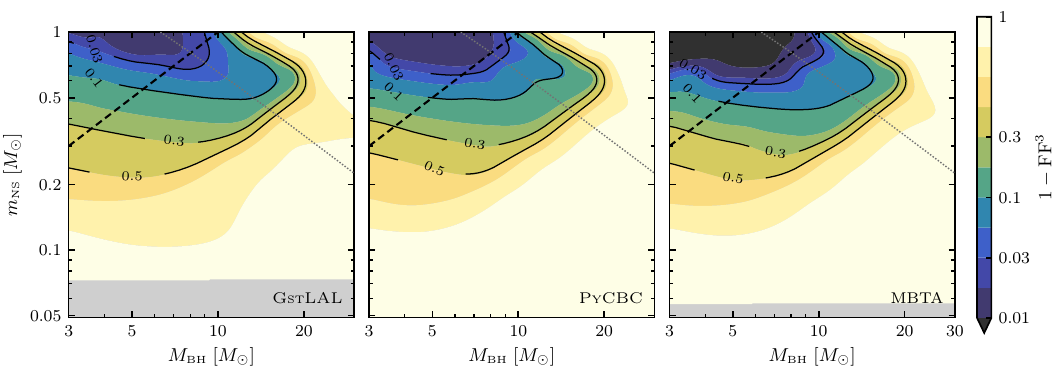}
\caption{Fraction of collapsar NS-BH events missed by the three O4 subsolar pipelines because of their templates, $1-{\rm FF}^3$, as in the left column of Fig.~\ref{fig:ssm_loss} ($f_\text{\tiny disk}=0$), for \textsc{GstLAL} ($45 - 1024\,$Hz), \textsc{PyCBC} ($30 - 1024\,$Hz), and MBTA ($35 - 512\,$Hz). Dashed: template-bank boundary, $q=0.1$. Dotted gray: $f_\text{\tiny Roche}=f_\text{\tiny ISCO}$. The gray region is the parameter space where the NS is disrupted before the lower cutoff of the pipeline.}
\label{fig:sm_pipelines_loss}
\end{figure*}

\begin{figure}[t]
\includegraphics[width=\linewidth]{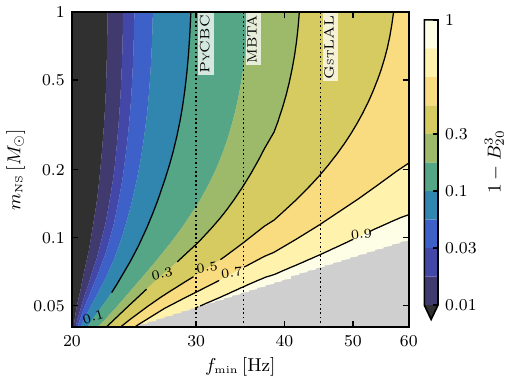}
\caption{Fraction of collapsar NS-BH events lost by starting the search at $f_\text{\tiny min}$ instead of $20\,$Hz, $1-B_{20}^3$ [Eq.~\eqref{eq:B20}], for our waveform with $M_\BH=5\,\msun$ and $f_\text{\tiny disk}=0$; it depends on $M_\BH$ only where the NS reaches the ISCO. The dotted lines identify the lower cutoffs of the three O4 pipelines. In the gray region, the NS is disrupted before $f_\text{\tiny min}$.}
\label{fig:sm_band_loss}
\end{figure}

\begin{figure*}[t]
\includegraphics[width=\linewidth]{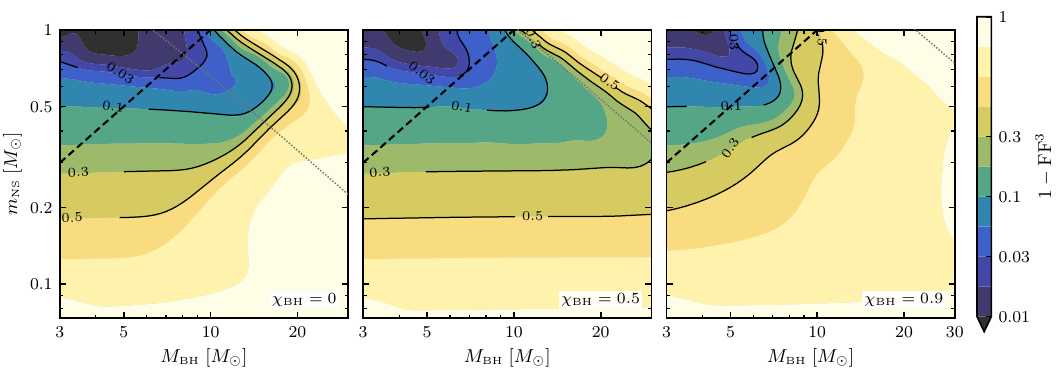}
\caption{Fraction of collapsar NS-BH events missed by the \textsc{GstLAL} search because of its templates, $1-{\rm FF}^3$, for BH spins $\chi_\BH=0$, $0.5$, and $0.9$ aligned with the orbital angular momentum, without the environmental phase ($f_\text{\tiny disk}=0$). Dashed: template-bank boundary, $q=0.1$. Dotted gray: $f_\text{\tiny Roche}=f_\text{\tiny ISCO}(\chi_\BH)$.}
\label{fig:sm_chibh}
\end{figure*}

\begin{figure*}[t]
\includegraphics[width=\linewidth]{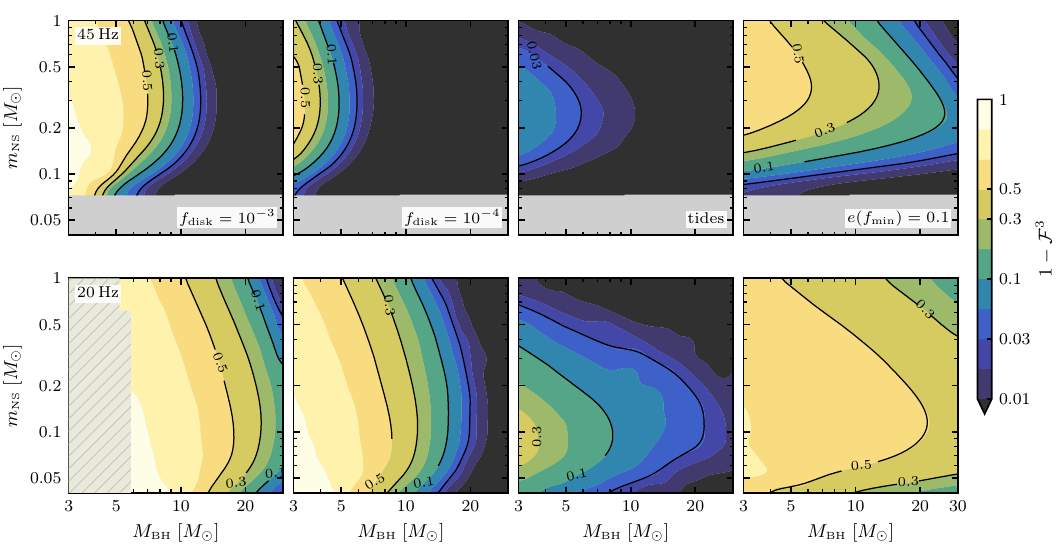}
\caption{Fraction of collapsar NS-BH events missed because of a single phase correction, $1-\mathcal{F}^3$ [Eq.~\eqref{eq:calF}], against aligned-spin \texttt{TaylorF2} templates that end at the disruption of the signal, with $q\geq10^{-4}$ and $|\chi_{1,2}|\leq0.99$. Top: $f_\text{\tiny min}=45\,$Hz; bottom: $f_\text{\tiny min}=20\,$Hz, both up to $1024\,$Hz. Columns, from left to right: migration with $f_\text{\tiny disk}=10^{-3}$ and $10^{-4}$, the tidal phase, and eccentricity $e_{\NS-\BH}=0.1$ at $f_\text{\tiny min}$. Gray: the NS is disrupted below $f_\text{\tiny min}$. Hatched: migration is faster than GW emission at $f_\text{\tiny min}$ [Eq.~\eqref{eq:fmig}], and Eq.~\eqref{eq:dpsi_env} does not apply. The SIQM term gives $1-\mathcal{F}^3\lesssim3\times10^{-2}$ in both bands and is not shown.}
\label{fig:sm_isolated}
\end{figure*}

\section{Template bank, pipelines, and black-hole spin}
\label{sec:pipelines}
\noindent
The LVK template bank consists of nonprecessing waveforms  and covers detector-frame masses $m_1\in[0.2,10]\,\msun$ and $m_2\in[0.2,1]\,\msun$, mass ratios $q\geq0.1$, and spin magnitudes $|\chi_{1,2}|\leq0.1$ for $m_{1,2}\leq0.5\,\msun$, and $|\chi_{1,2}|\leq0.9$ otherwise. The three search pipelines employ this common template bank with different frequency ranges, template-duration limits, and waveform approximants. \textsc{GstLAL} analyzes the $45 - 1024\,$Hz band, restricts template durations to $128\,$s, and uses \texttt{TaylorF2} for $\mathcal{M}_c\leq1.73\,\msun$ and \texttt{SEOBNRv4ROM}~\cite{Bohe:2016gbl} at larger chirp masses. \textsc{PyCBC} analyzes the $30 - 1024\,$Hz band, allows for template durations up to $1008\,$s, and uses \texttt{TaylorF2} for total masses $M\leq4\,\msun$ and \texttt{SEOBNRv4ROM} at larger masses. MBTA analyzes the $35 - 512\,$Hz band and uses \texttt{SpinTaylorT4} for $M<4\,\msun$ and \texttt{IMRPhenomTP} at larger masses; for aligned spins, we replace these approximants by \texttt{TaylorF2} and \texttt{IMRPhenomXAS}, respectively~\cite{Pratten:2020fqn}. In each case, the fitting-factor calculation uses the corresponding waveform approximant. The primary results are presented for \textsc{GstLAL}, which has the highest low-frequency cutoff, with comparisons to \textsc{PyCBC} and MBTA given below.

\emph{Other pipelines.}
Figure~\ref{fig:sm_pipelines_loss} compares the template losses $1-{\rm FF}^3$ of the three pipelines, each computed in its own band. They are comparable to those of \textsc{GstLAL}. The lower cutoffs of \textsc{PyCBC} and MBTA include more of the signal, but also more cycles over which the templates dephase, so the fitting factor does not increase and is slightly lower for light NSs. MBTA, whose band ends at $512\,$Hz (below the disruption frequency of the heaviest NSs), has the smallest template losses for heavy NSs. For light NSs, all three pipelines miss $\gtrsim90\%$ of the events.

We have also repeated the rate analysis for each pipeline, using its own O4a+O4b limits, interpolated as for \textsc{GstLAL}, and $\epsilon$ of Eq.~\eqref{eq:eps} computed in its own band. The efficiency $\epsilon$ is similar for the three pipelines: close to unity for heavy NSs and $\sim10^{-2}$ at $m_\NS\simeq0.1\,\msun$. The differences among the resulting limits therefore reflect mostly the published ones. Those of \textsc{PyCBC} are lower than those of \textsc{GstLAL} by a factor of $1.3 - 1.5$ in every bin, so its $\mathcal{R}_{90}/\epsilon$ falls below the LGRB rate over the widest region, $m_\NS\gtrsim0.5\,\msun$ and $M_\BH\simeq5 - 15\,\msun$, reaching $\simeq50\,{\rm Gpc^{-3}\,yr^{-1}}$ near $\mathcal{M}_c\simeq2\,\msun$; MBTA reaches the LGRB rate only in a narrow region near $M_\BH\simeq9\,\msun$. For light NSs, all three pipelines miss nearly every event and give comparable limits, $\sim10^5\,{\rm Gpc^{-3}\,yr^{-1}}$ at $m_\NS\simeq0.1\,\msun$.

\emph{Lower cutoff frequency.}
The template loss $1-{\rm FF}^3$ is computed relative to our waveform in the band of each pipeline. Relative to an ideal search that filters the same waveform from $20\,$Hz, a pipeline with low-frequency cutoff $f_\text{\tiny min}$ also loses the SNR accumulated below $f_\text{\tiny min}$, and detects only a fraction $B_{20}^3$ of the events, where
\begin{align}
    B_{20}&\equiv\frac{\rho_\sig(f_\text{\tiny min})}{\rho_\sig(20\,{\rm Hz})}\,,\label{eq:B20}\\
    \rho_\sig^2(f_\text{\tiny min})&=4\int_{f_\text{\tiny min}}\frac{|\tilde h_\sig(f)|^2}{S_n(f)}\,{\rm d}f\,,\nonumber
\end{align}
is the ratio of the optimal SNRs of the NS-BH waveform over the two bands, and $S_n$ is the O4 noise power spectral density. Like $B$ in Eq.~\eqref{eq:B}, $B_{20}$ is a property of the signal and not of the templates; unlike $B$, it compares the same collapsar signal in two bands, rather than the collapsar signal and a BH binary template in the same band. Figure~\ref{fig:sm_band_loss} shows that this loss is set by the NS mass, through $f_\text{\tiny Roche}\propto m_\NS$. Heavy NSs, disrupted far above $f_\text{\tiny min}$, lose about a third of the events at $45\,$Hz, the fraction of their SNR that the detector noise places between $20\,$Hz and $f_\text{\tiny min} = 45\,$Hz. Light NSs, disrupted only slightly above $f_\text{\tiny min}$, lose most of their in-band signal, and the loss grows steeply as $f_\text{\tiny min}$ approaches $f_\text{\tiny Roche}$: at $m_\NS\simeq0.1\,\msun$, it decreases from $\sim80\%$ for \textsc{GstLAL} to $\sim30\%$ for \textsc{PyCBC}. Unlike the template loss, this part of the loss would be recovered by lowering the cutoff of the search.

\emph{Black hole spin.}
The main text assumes non-spinning binaries. Figure~\ref{fig:sm_chibh} shows the template loss for a spinning BH for \textsc{GstLAL}. The spin moves the ISCO inward, so that the NS is tidally disrupted, rather than plunging, up to larger BH masses, and it adds spin-orbit and spin-spin terms to the phase, which accumulate over the many in-band cycles. For light NSs, the loss is set by the small mass ratio and the early disruption, and is nearly independent of $\chi_\BH$. For heavy NSs, it depends on $\chi_\BH$ non-monotonically. Templates at the bank boundary $q=0.1$ reproduce the phase of a $q\sim10^{-2}$ signal only with an offset primary spin, which grows faster than $\chi_\BH$. For $\chi_\BH=0$ this spin is strongly anti-aligned, as discussed in the main text. For $\chi_\BH=0.5$ it lies well inside the bank range, $|\chi_1|\leq0.9$, so that the losses remain at the level of tens of percent up to $M_\BH\simeq30\,\msun$, along the line $f_\text{\tiny Roche}=f_\text{\tiny ISCO}$. For $\chi_\BH=0.9$ the required spin exceeds the bank range, and the search misses most events with $M_\BH\gtrsim10\,\msun$, even for heavy NSs.

\section{Detection loss from individual phase corrections}
\label{sec:mismatch}
\noindent
The fitting factor ${\rm FF}$ of the main text combines several sources of mismatch: the small mass ratio of the signal, below the bank boundary $q=0.1$; its termination at tidal disruption, whereas the templates continue through merger; and the phase corrections of Eq.~\eqref{eq:dpsi_total}, which the templates do not contain. To separate the last from the others, we define a second fitting factor,
\begin{equation}
    \mathcal{F}(\bm{\theta})=\max_{\bm{\lambda},\,t_c,\,\phi_c}\frac{\langle h_\sig(\bm{\theta})|h_\mathcal{F}(\bm{\lambda})\rangle}{\sqrt{\langle h_\sig|h_\sig\rangle\,\langle h_\mathcal{F}|h_\mathcal{F}\rangle}}\,,
    \label{eq:calF}
\end{equation}
which has the form of Eq.~\eqref{eq:FF} but a different template family, 
\begin{equation}
    \tilde h_\mathcal{F}(f;\bm{\lambda})=\tilde h_\pp(f;\bm{\lambda})\,
    \mathcal{S}_*(f;\bm{\theta})\,\mathcal{S}_\text{\tiny end}(f;\bm{\theta})\,.
    \label{eq:hcalF}
\end{equation}
These templates are aligned-spin \texttt{TaylorF2} waveforms with the phase $\psi_\pp$ up to the 3.5PN order, multiplied by the high-frequency window of the signal, so that signal and template end at the same frequency. The template parameters $\bm{\lambda}$ span a broader domain than the search bank, with unrestricted chirp mass, $q\geq10^{-4}$, $|\chi_{1,2}|\leq0.99$, and no upper limit on template duration. The signal $\tilde h_\sig$ of Eq.~\eqref{eq:perturbed_waveform} contains a single phase correction in $\delta\psi$, with all other phase corrections set to zero. The inner product is taken over the $45 - 1024\,$Hz band of \textsc{GstLAL} and, to assess a lower cutoff, over $20 - 1024\,$Hz. The missed fraction $1-\mathcal{F}^3$ therefore quantifies the reduction in detection efficiency due solely to the selected phase correction. This calculation does not correspond to an existing search, but it provides a lower bound on the loss for any aligned-spin template bank contained within the template family defined above.

The extent to which a phase correction can be reproduced by shifts in the template parameters depends primarily on its frequency dependence. Relative to the Newtonian phase $\psi_\text{\tiny N}\propto v^{-5}$, the template phase contains terms proportional to $v^0$ through $v^7$, with coefficients that depend on the chirp mass at Newtonian order, the symmetric mass ratio from 1PN order, and the aligned spins from 1.5PN order. The coalescence time and phase contribute terms proportional to $v^8$ and $v^5$, respectively. Over a finite frequency interval, variations in these parameters can approximately reproduce a correction whose relative power of $v$ lies within this range, whereas corrections with powers well outside it are generally less degenerate with the template parameters. We discuss the four corrections in order of decreasing impact on detection efficiency; Fig.~\ref{fig:sm_isolated} shows those associated with a non-negligible fraction of missed events.

The migration term ($-4$PN) is the largest correction whenever residual gas is present. Its size relative to the GW-driven phase, $\propto(Mf)^{-8/3}$, increases toward low frequencies and light BHs and, at fixed frequency, depends only on $M\simeq M_\BH$; accordingly, the loss depends weakly on $m_\NS$. Since a relative $v^{-8}$ correction lies outside the range of powers in the template phase, the templates reproduce it only partially, through a larger chirp mass, near-extremal spins, and a smaller mass ratio. In the $45\,$Hz band, the loss is large for light BHs when $f_\text{\tiny disk}=10^{-3}$, and limited to the lightest BHs when $f_\text{\tiny disk}=10^{-4}$. For heavy NSs this loss matches the extra bank loss in the second column of Fig.~\ref{fig:ssm_loss}.

The tidal phase (5PN) is largely absorbed by a smaller $q$, because over the short band between $f_\text{\tiny min}$ and disruption a shift of $q$, together with $\mathcal{M}_c$ and the spins, approximates its frequency dependence. The residual loss, $\lesssim10\%$ at $45\,$Hz, is largest for intermediate-mass NSs around light BHs, for which the PN phase $\propto q^4\Lambda$ is largest and the disruption frequency is close to the minimum of the noise spectrum.

The eccentric term $(-19/6$PN) scales as $e^2$ [Eq.~\eqref{eq:dpsi_ecc}] and is largest at the lower cutoff of the band. For $e_\text{\tiny NS-BH}=0.1$ at $f_\text{\tiny min}$ (Fig.~\ref{fig:sm_isolated}), shifts of the template parameters do not reproduce this correction, and the loss is largest for light BHs, whose signals contain the most cycles. Because GW emission circularizes the orbit ($e\propto f^{-19/18}$ for $e\ll1$), an eccentricity of $0.1$ at $20\,$Hz decreases by more than a factor of two at $45\,$Hz; for this value of eccentricity in the $45\,$Hz band, small shifts of $\mathcal{M}_c$, $q$, and $\chi_1$ then reproduce the correction, and the loss is a few percent (not shown). The SIQM term (2PN) is absorbed by a chirp-mass shift of order $10^{-3}$, even for $\chi_\NS=1$, because it has the same frequency dependence as the 2PN spin-spin terms of the templates; the loss is at the percent level.

\emph{Lower cutoff frequency.}
Lowering $f_\text{\tiny min}$ to $20\,$Hz has two opposite effects on detection. It increases the optimal SNR, since the long inspiral accumulates a significant fraction of its SNR between $20\,$Hz and $45\,$Hz (Fig.~\ref{fig:sm_band_loss}), and it brings into the band the lightest NSs, which are disrupted below $45\,$Hz. It also increases the number of in-band cycles, approximately as $f_\text{\tiny min}^{-5/3}$ for signals that end well above $f_\text{\tiny min}$, and adds the frequencies at which the negative-PN corrections are largest. Over the longer band the PN velocity spans a wider range, and the template parameters, which are fixed by the phase evolution over the whole band, reproduce a correction with a different frequency dependence less accurately. The template losses therefore increase for all corrections (bottom row of Fig.~\ref{fig:sm_isolated}); the SIQM loss increases slightly and remains at the percent level. Migration at $f_\text{\tiny disk}=10^{-4}$ causes losses comparable to those at $f_\text{\tiny disk}=10^{-3}$ in the $45\,$Hz band, and at $f_\text{\tiny disk}=10^{-3}$ the migration-driven evolution exceeds the GW-driven one at band entry for the lightest BHs [Eq.~\eqref{eq:fmig}]. The residual tidal loss increases, mostly for the light NSs that now enter the band well before disruption and that have large tidal deformabilities. An eccentricity of $0.1$ at $20\,$Hz, reproduced by the templates in the $45\,$Hz band, causes a large loss over the whole parameter space.

The SNR gain from a lower cutoff is therefore partly offset by larger template losses unless the templates contain these corrections. With such templates, the low-frequency cycles that reduce the template match carry most of the information on migration and eccentricity, and thereby distinguish collapsar sources from PBH binaries. The migration correction requires dedicated templates in both bands; the tidal and eccentric phase corrections also require dedicated templates in the $20\,$Hz band.

\end{document}